\documentclass[%
 aip,
 jmp,%
 amsmath,amssymb,
preprint,
]{revtex4-1}

\usepackage{graphicx}
\usepackage{dcolumn}
\usepackage{bm}
\usepackage[version=3]{mhchem}
\usepackage{threeparttable}
\usepackage[usenames]{color}
\usepackage{footnote}
\usepackage{multirow}
\usepackage{url}

\newcommand{\subm}[1]{{\textit{\tiny #1}}} 
\newcommand{\subt}[1]{{\textit{#1}}} 

\begin{document}

%
\title[]{A quantum-mechanical perspective on linear response theory within polarizable embedding}

\author{Nanna Holmgaard List}
\thanks{Author to whom correspondence should be addressed. Electronic mail: nalist@kth.se}
\affiliation{Division of Theoretical Chemistry and Biology, School of Biotechnology, KTH Royal Institute of Technology, Roslagstullsbacken 15, SE-106 91 Stockholm, Sweden}
\author{Patrick Norman}
\email{panor@kth.se}
\affiliation{Division of Theoretical Chemistry and Biology, School of Biotechnology, KTH Royal Institute of Technology, Roslagstullsbacken 15, SE-106 91 Stockholm, Sweden}
\author{Jacob Kongsted}
\email{kongsted@sdu.dk}
\affiliation{Department of Physics, Chemistry and Pharmacy, University of Southern Denmark, Campusvej 55, 5230 Odense M, Denmark}
\author{Hans J\o{}rgen Aagaard Jensen}
\email{hjj@sdu.dk}
\affiliation{Department of Physics, Chemistry and Pharmacy, University of Southern Denmark, Campusvej 55, 5230 Odense M, Denmark}
%
%

\begin{abstract}
The derivation of linear response theory within polarizable embedding is carried out from a rigorous quantum-mechanical treatment of a composite system. Two different subsystem decompositions (symmetric and nonsymmetric) of the linear response function are presented, and the pole structures as well as residues of the individual terms are analyzed and discussed. This theoretical analysis clarifies 
which form of the response function to use in polarizable embedding,
and we highlight complications in separating out subsystem contributions to molecular properties. For example, based on the nonsymmetric decomposition of the complex linear response function, we derive conservation laws for integrated absorption cross sections, providing a solid basis for 
proper calculations of the intersubsystem intensity borrowing inherent to coupled subsystems
and how that can lead to negative subsystem intensities.
We finally identify steps and approximations required to achieve the transition from a quantum-mechanical description of the composite system to polarizable embedding with a classical treatment of the environment, thus providing a thorough justification for the descriptions used in polarizable embedding models.

\end{abstract}

\pacs{33.20.-t, 33.70.Ca, 33.80.-b}
\keywords{polarizable embedding, response theory, localized electronic transition, subsystem decompositions, effective external field effect}
\maketitle

\section{Introduction}\label{intro}
The many types of spectroscopy that have been developed over the years (some of which are still under development) serve as an indispensable tool to gain fundamental understanding of the structure and dynamics of molecular systems.\cite{Lindon2010} Equally indispensable are the large number of theoretical methods that nowadays allow for accurate \textit{in silico} simulations of said systems and spectroscopies, and which can provide complementary information that is inaccessible in the experiment.
When designing calculations of molecular properties, it is often necessary to consider the influence of the surrounding environment to obtain adequate accuracy. Given the typical steep computational scaling of quantum-chemical methods with respect to the number of basis functions, the description of large molecular systems is out of question using conventional algorithms. 

One way to partially address the challenges of treating large systems is to exploit localization techniques and simplified treatments of long-range interactions to achieve reduced, ultimately linear scaling.\cite{kristensen2012mp2,ziolkowski2010linear,ochsenfeld2007linear,sherrill2010frontiers} While linear scaling formulations, aided by increasing computer power, significantly extend the applicability range of quantum chemistry, they are generally too costly for routine applications, especially if the system and property of interest put special demands on method, basis set and conformational sampling. For instance, even in cases where (time-dependent) density functional theory ((TD)DFT) is still feasible, it may suffer from electron-transfer overstabilization in the ground state \cite{jakobsen2013electrostatic,olsen2015accuracy} and artificial low-lying charge-transfer transitions\cite{dreuw2003long,gritsenko2004asymptotic,dreuw2004failure} due to the self-interaction errors associated with approximate standard exchange--correlation functionals.
Another complication is that the delocalized picture obtained from such brute-force calculations is not straightforward to interpret in terms of chemical concepts and generally requires additional decomposition analyses of the wave function and properties into local components. 

An alternative route is to adopt subsystem approaches in which the system is divided into chemically well-defined constituents which may be treated as separate entities perturbed by the other subsystems, and as such they offer a natural decomposition into environmental effects. Several methods belong to this category and one may generally distinguish two conceptually different approaches: (\textit{i}) strict subsystem methods that treat the subsystems consistently on an equal footing and which upon recombination yield the properties of the entire system,\cite{gordon2012fragmentation,jacob2014subsystem} and (\textit{ii}) the so-called embedding approaches in which an accurate quantum-mechanical description is intended for only a smaller part of the system, while the remaining parts (the environment) and its effects on the former are described with more efficient, approximate methods. A recent review of subsystem and embedding approaches can be found in Ref.~\citenum{gomes2012quantum}. Apart from the obvious computational cost argument, embedding approaches are motivated by the fact that the chemical identity of subsystems is largely intact in molecular complexes and that many electronic transitions are localized in nature.
Prototypical examples are solute--solvent systems and chromophore--protein complexes. 
Frozen-density embedding\cite{wesolowski1993frozen} (FDE) provides in principle a full quantum-mechanical description of the system based on a subsystem formulation of DFT and can be applied in both a self-consistent and an embedding mode thereby bridging (\textit{i}) and (\textit{ii}). In principle, it is a formally exact framework although its quality with respect to supermolecular DFT in practical calculations is limited by approximations in the employed nonadditive parts of the exchange--correlation and kinetic energy functionals.

More efficient, though more approximate embedding approaches are quantum--classical models, such as hybrid quantum mechanics/molecular mechanics methods, which use a discrete but classical representation of the environment.\cite{warshel1976theoretical,senn2009qm,van2013combined} These approaches are classified in a hierarchy according to the extent of the coupling between the quantum and classical subsystems. In electrostatic embedding, the environment is represented as a purely external electrostatic potential perturbing the density of the quantum subsystem, while a more realistic description incorporating the mutual polarization effects between the quantum and classical subsystems is offered by polarizable embedding schemes. A variety of embedding schemes to model environment polarization have been proposed and applied, such as induced dipoles and fluctuating charges.\cite{lopes2009molecular} 
Inclusion of the explicit environmental response is particularly important for processes where substantial rearrangements may occur, e.g., upon electronic excitation.\cite{sneskov2011scrutinizing} 
In the FDE scheme, state-specific polarization of the environment can be included in the self-consistent mode of the formalism by interchanging the roles of the subsystems in so-called freeze-and-thaw iterations.\cite{wesolowski1996kohn}
In polarizable embedding, this is replaced by a self-consistent determination of the model-specific embedding parameters describing the polarization of the environment.

Both FDE\cite{casida2004generalization,hofener2012molecular,neugebauer2010chromophore,neugebauer2009calculation,neugebauer2007couplings} and polarizable embedding approaches\cite{olsen2010,lipparini2012linear,EFP-excited-states,list2016excited,jensen2003discrete2,morton2011discrete,curutchet2009electronic} have been generalized to a response formalism to allow for the calculation of response and transition properties of molecules embedded in large molecular complexes. The computational cost associated with the explicit coupling of the subsystem excitation manifolds in a fully coupled FDE scheme generally hinders the inclusion of the dynamical response of the entire environment in large complexes,\cite{neugebauer2009calculation} but allows in a truncated form to describe a few strongly coupled local excitations as relevant for chromophoric aggregates.\cite{neugebauer2007couplings} The classical treatment in polarizable embedding schemes, on the other hand, offers an efficient inclusion of the dynamic environmental response, provided the perturbing field is nonresonant with respect to local excitations in the environment.\cite{jacob2006comparison} This restriction has, however, been lifted by introducing phenomenological excited-state lifetimes in the response formalism, see Refs.~\citenum{morton2011discrete,payton2012discrete}. 

The objective of the present work is to formulate a rigorous derivation of linear response theory within polarizable embedding starting from a quantum-mechanical treatment of the entire system. While various polarizable embedding schemes differ in the specific representation of the environment, the underlying mathematical structure and physical content of their working equations are the same.\cite{list2013unified} For the present aim, we will focus on the explicit expressions for the polarizable embedding (PE) model\cite{olsen2010,olsen2011,list2016excited} in which the environment is described in terms of a distributed multipole representation, thus providing a well-defined link to the environment charge density. We note that the main conclusions will also be valid for, e.g., polarizable density embedding,\cite{olsen2015polarizable} which goes beyond pure polarizable embedding.
While the formally exact FDE framework has previously been used as a common theoretical platform for discussing embedding models,\cite{gomes2012quantum} our derivation will be based on an explicit parameterization of the wave function of the combined system. The origin and inherent limitations of polarizable embedding become particularly transparent in this framework since approximations are introduced by imposing restrictions directly on the corresponding wave function parameterization, as we shall see.

The implications of the choice of parameterization are pertinent to the extension to response properties: the equivalence between response theory and state-specific approaches is lost when employing a nonlinear parameterization,\cite{olsen1995} as is the case for polarizable embedding, such that different physical pictures emerge in the two formalisms. 
On the basis of a four-state model, it was shown in Ref.~\citenum{corni2005electronic1} that the exact second-order excitation energy for a system of two interacting subsystems contains three contributions in addition to the zeroth-order term describing the excitation within a frozen environment: (\textit{i}) a term describing the differential polarization in the environment upon excitation in the embedded molecule, i.e., a classical induction effect, (\textit{ii}) the difference in dispersion interaction between the environment and the embedded molecule in its ground state and excited state as given by the Casimir--Polder formula, and (\textit{iii}) a term that describes the coupling between the excited state of interest with all lower-lying states of the embedded molecule, i.e., de-excitations, in terms of a screened interaction mediated by the environment through its dynamic polarizability (evaluated at the excitation frequency). For a generalization beyond a few-state model, see Refs.~\citenum{list2015thesis,schwabe2016general}. Regardless the debated interpretation of this term as  
a dispersion effect,\cite{corni2005electronic1,roesch1994calculation,mcrae1957theory,buhmann2012dispersion} a nonresonant excitonic coupling between subsystems\cite{lunkenheimer2012solvent,schwabe2016general} or a resonant coupling,\cite{daday2015chromophore} it is a manifestation of the mutual correlation of the electrons in the subsystems.  
Employing a direct-product \textit{ansatz} in a linear response framework recovers only the third term, although it should be noted that the correspondence is incomplete for all but the lowest excitation since de-excitations to lower-lying states different from the ground state are missing in the response treatment based on a ground-state reference.\cite{list2015thesis} On the other hand, only the state-specific induction term (\textit{i}) is recovered in a state-specific formalism. 

Whereas previous theoretical analyses have primarily focused on excitation energies,\cite{gomes2012quantum,schwabe2016general,corni2005electronic1} we shall here be concerned with the general linear response function, and not only its poles. Based on the linear response function of the combined system, we will derive two different subsystem decompositions, referred to as symmetric and nonsymmetric, respectively. These provide a direct link to the standard response formulations of polarizable embedding based on the classical description of the energy contribution of the interaction with the environment.\cite{jensen2005microscopic,list2016excited}
In addition to justifying the form of the environmental effects in polarizable embedding schemes, the present analysis also sheds light on the basic features of coupled systems and potential complications in separating out subsystem contributions to molecular properties. As discussed by Pavanello in the subsystem DFT framework, transitions in composite systems are not strictly localized in the sense that the pole structure of the combined system is inherited by the individual subsystem contributions to the linear response function.\cite{pavanello2013subsystem} We will extend that analysis by considering the consequences on the evaluation of transition strengths in an embedding framework by contrasting the symmetric and nonsymmetric decompositions. 
This exposition will further benefit from our consideration of the complex linear response function of the combined system that puts emphasis on conservation laws for integrated intensities and their ramifications for the decomposed forms of the response function used in practical polarizable embedding schemes.

Finally, the key features of linear response theory of a combined system subjected to a weak external electric field will be illustrated numerically by considering a six-level-model of a hydrogen-bonded dimer complex formed by water and \textit{para}-nitroaniline. This simple example serves to demonstrate the results otherwise found in the theoretical analysis.  

\section{Theory}\label{sec:theory}
Our presentation is intended to be self-contained and thus we provide a rather extensive theory section, starting in the first part with a brief review of the conventional quantum-mechanical direct-product  \textit{ansatz} for the wave function of the combined system.\cite{angyan1992common,mcweeny1969methods,corni2005electronic1}
We then provide a rigorous formulation of linear response theory for a composite system within this theoretical framework. 
Our formulation allows, upon additional approximations, to recover the PE model. In practice, we exploit two different decompositions that allow for solving the response equations in effective subsystem spaces. This leads to the definition of effective quantities that provide insight into how subsystem properties are modified in the presence of another interacting subsystem.
The features of the individual terms in the two decompositions are compared with a view to the evaluation of subsystem contributions to response and transition properties. Finally, to overcome some of the challenges related to the subsystem formulation in conventional linear response theory, we extend the treatment to a complex response framework. In addition, such formulation allows to demonstrate the intensity borrowing occurring in a coupled system.  

In the second part of the this section, we analyze the problem of recovering the PE model and its extension to quantum-mechanical linear response theory based on the decompositions derived in the first part. Specifically, we assume a perturbation treatment of the environment subsystem as well as a multipole representation of its charge distribution. In this way, we provide a rigorous derivation of the environmental effects emerging in linear response theory within polarizable embedding.\cite{list2016excited}

Atomic units (a.u.) will be used throughout and, in order to keep equations more compact, we will allow ourselves to leave out physical constants that have a numerical value of 1 a.u.

\subsection{Preliminary Considerations}
Consider a composite system consisting of $N$ interacting subsystems each with an integer number of electrons and with a fixed relative position. A fundamental assumption made in quantum--classical embedding models is that of nonoverlapping subsystem charge densities (zero-overlap approximation), which implies that exchange--repulsion vanishes. 
As a consequence, the exact wave function of the combined system can be written in the basis of direct-product states constructed from the complete antisymmetrized subsystem spaces.\cite{mcweeny1969methods,corni2005electronic1} 
The nonrelativistic Born--Oppenheimer Hamiltonian
for the interacting system decomposes naturally as
\begin{align}
\hat{\mathcal{H}}=\sum_{\subm{A}=1}^N\hat{\mathcal{H}}_{\subm{A}}+\sum_{\subm{A}> \subm{B}}^N\hat{\mathcal{V}}_{\subm{AB}}\ \text{,} 
\end{align}
where $\hat{\mathcal{H}}_{\subm{A}}$ is the electronic Hamiltonian of the isolated subsystem \subt{A} and 
$\hat{\mathcal{V}}_{\subm{AB}}$ the interaction operator describing the interactions between the nuclei and 
electrons in subsystem \subt{A} with those in \subt{B}. 
In second quantization, the interaction operator takes the form
\begin{align}\label{eq:interaction_operator}
\hat{\mathcal{V}}_{\subm{AB}} =&\sum_{m\in \subm{B}}^{M_{\subm{B}}}Z_m\sum_{pq\in \subm{A}}v_{pq}(\mathbf{R}_m)\hat{E}_{pq}+\sum_{n\in \subm{A}}^{M_{\subm{A}}}Z_n\sum_{rs\in \subm{B} }v_{rs}(\mathbf{R}_n)\hat{E}_{rs}\notag\\ 
&+\sum_{pq\in \subm{A}}\sum_{rs \in \subm{B}} v_{pq,rs}\hat{E}_{pq}\hat{E}_{rs}+ \sum_{n\in \subm{A}}^{M_{\subm{A}}}\sum_{m\in \subm{B}}^{M_{\subm{B}}}\frac{Z_nZ_m}{|\mathbf{R}_n-\mathbf{R}_m|}\ \text{,}
\end{align}
where $Z_n$ and $\mathbf{R}_n$ denote the charge and the position vector, respectively, of nucleus $n$ of the $M_{\subm{A}}$ nuclei in subsystem \subt{A}, while $Z_m$, $\mathbf{R}_m$ and $M_{\subm{B}}$ are the corresponding quantities belonging to subsystem \subt{B}.
$\hat{E}_{pq}$ is the usual second quantization singlet one-electron excitation operator,\cite{helgaker2000molecular} and $p,q,r,s$ are
used as general spatial molecular orbital indices. The affiliation of the corresponding orbitals
 to a given subsystem is indicated by the summation.  
The associated one- and two-electron integrals are defined as
\begin{align}
v_{pq}(\mathbf{R}) &= -\int \frac{\phi_p^*(\mathbf{r})\phi_q(\mathbf{r})}{|\mathbf{R}-\mathbf{r}|} \text{d}\mathbf{r}\ \text{,}\\[0.1in] 
v_{pq,rs} &= \iint \frac{\phi_p^*(\mathbf{r}_1) \phi_r^*(\mathbf{r}_2) \phi_q(\mathbf{r}_1)  \phi_s(\mathbf{r}_2)}{|\mathbf{r}_1-\mathbf{r}_2|}\, \text{d}\mathbf{r}_1\, \text{d}\mathbf{r}_2\ \notag \\
&= -\int \phi_p^*(\mathbf{r}_1) \phi_q(\mathbf{r}_1) v_{rs}(\mathbf{r}_1)\  \text{d}\mathbf{r}_1\  \text{.}\ \label{eq:2e_integral}
\end{align}
The first two terms in Eq.~\eqref{eq:interaction_operator} 
describe the instantaneous intersubsystem Coulomb interaction between the electrons in 
subsystem \subt{A} and the $M_{\subm{B}}$ nuclei in subsystem \subt{B} and vice versa, whereas the 
third and fourth terms are the 
intersubsystem 
electron--electron and nucleus--nucleus repulsion terms, respectively. Notice that 
no exchange integrals between the subsystems survive within the zero-overlap approximation, and the two-electron excitation operator therefore factorizes into subsystem contributions. 
Occasionally, we will use an alternative representation of the interaction operator
\begin{align}\label{eq:potential_operator}
\hat{\mathcal{V}}_{\subm{AB}}=\int\hat{\rho}_\subm{A}(\mathbf{r})\hat{\mathcal{V}}_\subm{B}(\mathbf{r})\text{d}\mathbf{r}=\int\hat{\rho}_\subm{B}(\mathbf{r})\hat{\mathcal{V}}_\subm{A}(\mathbf{r})\text{d}\mathbf{r}=\hat{\rho}^\subm{A}_\mathbf{r}\ \hat{\mathcal{V}}^\subm{B}_\mathbf{r}\ \text{,}
\end{align}
given in terms of the first-order reduced density and electrostatic potential operators:
\begin{align}
\hat{\rho}_{\subm{A}}(\mathbf{r})=&
\sum_{n\in \subm{A}}^{M_{\subm{A}}}Z_n\delta(\mathbf{r}-\mathbf{R}_n)-\sum_{pq\in \subm{A}}\phi_p^*(\mathbf{r})\phi_q(\mathbf{r})\hat{E}_{pq}~,\label{eq:den_operator}\\
\hat{\mathcal{V}}_{\subm{B}}(\mathbf{r})=&
\sum_{m\in \subm{B}}^{M_{\subm{B}}}\frac{Z_m}{|\mathbf{r}-\mathbf{R}_m|}+\sum_{rs\in \subm{B}}v_{rs}(\mathbf{r})\hat{E}_{rs}\ \text{.}\label{eq:pot_operator}\displaybreak[1]
\end{align}
The last equality of Eq.~\eqref{eq:potential_operator} introduces a shorthand notation for the spatial integration with respect to repeated space variables.

To maximize the comparability with the effective environment models, we approximate the electronic wave function of the combined system with a single direct product of subsystem wave functions 
\begin{align}
\label{eq:wavefunction}
|\Psi_{\subm{A}}\Psi_{\subm{B}} \dots\Psi_{\subm{N}}\rangle
 &= |\Psi_{\subm{A}}\rangle\otimes |\Psi_{\subm{B}}\rangle\otimes \dots|\Psi_{\subm{N}}\rangle\ \text{.}
\end{align}
The variation of the associated energy functional according to the Rayleigh--Ritz variational principle leads to a set of coupled effective subsystem equations
\begin{align}\label{eq:effective_equations}
\forall\  \subt{A}:\hspace{10pt}\hat{\mathcal{F}}_\subm{A}|0_{\subm{A}}\rangle=E_\subm{A}|0_{\subm{A}}\rangle \ ,
\end{align}
where the effective subsystem Hamiltonian is given by
\begin{align}
\hat{\mathcal{F}}_\subm{A}&=\hat{\mathcal{H}}_{\subm{A}}+\sum_{\subm{B}\neq\subm{A}}^N\langle 0_{\subm{B}}| \hat{\mathcal{V}}_{\subm{AB}}| 0_{\subm{B}}\rangle\notag\\
&=
 \hat{\mathcal{H}}_{\subm{A}} + \sum_{pq\in {\subm{A}}}\sum_{{\subm{B}\neq\subm{A}}}
 \biggl[\sum_{m\in {\subm{B}}}^{M_{\subm{B}}}Z_mv_{pq}(\mathbf{R}_m)+\sum_{rs\in {\subm{B}}}v_{pq,rs}D_{\subm{B},rs}\biggl]\hat{E}_{pq}\ \text{.}\label{eq:eff_Hamiltonian}
\end{align}
Here, $D_{\subm{B},rs}=\langle 0_\subm{B} | \hat{E}_{rs}| 0_\subm{B}\rangle$ is an element of the first-order reduced density matrix for subsystem \subt{B}. The two terms collected in the square bracket represent the classical electrostatic potential generated by the ground states of the remaining subsystems in their polarized states, i.e., in the presence of the other subsystems.  In practice, any wave function \textit{ansatz} may be invoked for the individual subsystems and the associated optimization conditions derived upon applying the variational principle. Equation~\eqref{eq:eff_Hamiltonian} is analogous to Hartree self-consistent field theory, where the orbitals replace the subsystem wave functions. In particular, the parameterization in Eq.~\eqref{eq:wavefunction} discards direct-product states in which the subsystems are simultaneously excited,
which implies that no dispersion effects are included in this approximation. 
However, some intersubsystem electron correlation effects are introduced 
when the direct-product \textit{ansatz} is used in a response framework, as mentioned in the Introduction. 

\section{Response theory of composite systems}
We will now consider the electronic response of the composite system to an external optical field within the framework introduced in the previous section.
In such a case, nuclear motions are neglected and only the pure electronic response will be considered. 
Special attention will be paid to the physical aspects of the intersubsystem interactions in the presence of the applied field, and how they influence molecular properties.
For the sake of notational simplicity, we shall restrict this analysis to two subsystems (leading to an embedded subsystem \subt{A} and an environment \subt{B} in polarizable embedding) and work within a configuration interaction (CI) framework for the individual subsystems. The framework can easily be extended to consider the individual subsystems of the environment  still assuming nonoverlapping subsystems, by decomposing the environment wave function into a product of subsystem contributions (see Eq.~\ref{eq:wavefunction}). This will be used in Sec.~\ref{sec:PE}. We also note that, even if we are considering explicitly the CI parameterization for wave functions of the individual subsystems, the principles apply more generally also to other variational wave function models.

The field--matter interaction will be described within the semi-classical framework, 
where the incident field 
is treated as a classical plane wave that perturbs  the molecular system. 
The perturbation operator describing the action of a monochromatic external field on the composite system can then be expressed in terms of Fourier components as
\begin{align}
\hat{V}^t&= \sum_{\pm\omega}\hat{V}_{\alpha}^{\omega}F_{\alpha}^{\omega}e ^{-i\omega t}\notag\\
&=\sum_{\pm\omega}\left(\hat{V}_{\subm{A},\alpha}^{\omega}+\hat{V}_{\subm{B},\alpha}^{\omega}\right)F_{\alpha}^{\omega}e ^{-i\omega t}\ \text{,}\label{eq_pertop}
\end{align}
where $F_{\alpha}^{\omega}$ are Fourier amplitudes 
associated with the one-electron perturbation operators $\hat{V}_{\alpha}^{\omega}$, using Greek subscripts as (possibly composite) Cartesian labels. In order to maintain Hermiticity of $\hat{V}^t$, we have $F_{\alpha}^{\omega}=[F_{\alpha}^{-\omega}]^*$ and 
$\hat{V}_{\alpha}^{\omega}=[\hat{V}_{\alpha}^{-\omega}]^{\dagger}$.
Note that the additivity of the perturbation operator in the last equality of Eq.~\eqref{eq_pertop} is a consequence of the zero-overlap assumption.
Here and henceforth, the Einstein summation convention is adopted for repeated Greek indices.

Our derivation will follow the quasi-energy formulation of response theory,\cite{christiansen1998response,sasagane1993} which can be viewed as a time-dependent generalization of the ordinary energy-differentiation technique from time-independent perturbation theory, which in its time-averaged formulation is restricted to time-periodic perturbations. 
The response functions are defined as the coefficients of the time-averaged quasi-energy in a Taylor series
expansion in terms of the external field strengths
\begin{align}\label{eq:TA_quasi_energy}
\{Q_{\subm{AB}}\}_T &= E_{\subm{AB}} + \sum_{\omega}\langle 0_{\subm{A}}0_{\subm{B}}|\hat{V}_{\alpha}^{\omega}|0_{\subm{A}}0_{\subm{B}}\rangle F^{\omega}_{\alpha}\delta_{\omega}\notag\\[0.1in]
&+\frac{1}{2}\sum_{\omega_1,\omega_2}\langle \langle \hat{V}_{\alpha}^{\omega_1};\hat{V}_{\beta}^{\omega_2}\rangle\rangle\delta_{\omega_1+\omega_2} F_{\alpha}^{\omega_1}F_{\beta}^{\omega_2}\notag\\[0.1in]
&+\frac{1}{6}\sum_{\omega_1,\omega_2}\langle \langle \hat{V}_{\alpha}^{\omega_1};\hat{V}_{\beta}^{\omega_2},\hat{V}_{\gamma}^{\omega_2}\rangle\rangle\delta_{\omega_1+\omega_2+\omega_3} F_{\alpha}^{\omega_1}F_{\beta}^{\omega_2}F_{\gamma}^{\omega_2}+\cdots \ \text{,}
\end{align}
where $\{\cdot\}_T$ indicates that the time-average over one period of oscillation in the external field has been taken. The symbol $\delta_\omega$, not to be confused with a Dirac-delta or a discrete Kronecker-delta function, is unity when the continuous frequency variable vanishes and zero otherwise.
An instructive exposition of the quasi-energy derivative approach as well as a comparison to the alternative Ehrenfest formulation can be found in Ref.~\citenum{norman2011perspective}.

\subsection{The Direct-Product Approximation}\label{sec:compositesystems}
The phase-isolated part of the time-dependent direct-product wave function for the composite system may be defined 
by an exponential unitary parameterization as
\begin{align}\label{eq:LRCS_parameterization}
|\widetilde{0_{\subm{A}} 0_{\subm{B}}}\rangle &= e^{{i}\left[\hat{\Lambda}_{\subm{A}}(t)+\hat{\Lambda}_{\subm{B}}(t)\right]}|0_{\subm{A}}0_{\subm{B}}\rangle\notag\\
&=e^{{i}\hat{\Lambda}_{\subm{A}}(t)}|0_{\subm{A}}\rangle\otimes e^{{i}\hat{\Lambda}_{\subm{B}}(t)}|0_{\subm{B}}\rangle; \hspace{20pt}[\hat{\Lambda}_{\subm{A}},\hat{\Lambda}_{\subm{B}}]=0\ \text{.}
\end{align}
The intersubsystem commutation relation above follows from the zero-overlap assumption.
The time-dependent Hermitian operator $\hat{\Lambda}_{\subm{A}}(t)$ for subsystem \subt{A} is parameterized 
in terms of a set of time-dependent amplitudes $\{\lambda_{i_\subm{A}}\}$ and takes the form\cite{olsen1985linear}
\begin{align}
\hat{\Lambda}_{\subm{A}}(t) = \sum_{i> 0 } \left(\lambda_{i_{\subm{A}}}(t)\hat{q}_{i_{\subm{A}}}^{\dagger}+\lambda^*_{i_{\subm{A}}}(t)\hat{q}_{i_{\subm{A}}} \right) =\mathbf{Q}_\subm{A}^\dagger\bm{\Lambda}_\subm{A}\ \text{,}\label{eq:operator}\\
\mathbf{Q}_{\subm{A}}^{\dagger}=\begin{bmatrix}
\mathbf{q}_\subm{A}^{\dagger}& \mathbf{q}_\subm{A}\end{bmatrix};\hspace{10pt} \mathbf{\Lambda}_\subm{A}=\begin{bmatrix}\bm{\lambda}_\subm{A}(t) & \bm{\lambda}_\subm{A}(t)^*\end{bmatrix}^T\ \text{,}\notag
\end{align}
where an identity operator for subsystem \subt{B} is implied.
The state-transfer operators
$\hat{q}_{i_{\subm{A}}}^{\dagger}{=}|i_{\subm{A}}\rangle\langle 0_{\subm{A}}|$ and their adjoints 
are built from a set of orthonormalized states $\{|i_{\subm{A}}\rangle\}$ 
that spans the orthogonal complement space of the reference state $|0_{\subm{A}}\rangle$, 
the latter satisfying the variational condition given by
Eq.~\eqref{eq:effective_equations}. Before proceeding, we briefly comment on the employed parameterization: (\textit{i}) The direct-product parameterization in Eq.~\eqref{eq:LRCS_parameterization} is nonlinear: that is, it produces states outside the excitation manifold, defined by the linear action of the operator $\hat{\Lambda}_{\subm{A}}(t)+\hat{\Lambda}_{\subm{B}}(t)$. In particular, the exponential parameterization contains states in which both subsystems are excited simultaneously due to products of subsystem excitations, despite the absence of such transitions in the state-transfer operators. This is analogous to the nonlinear exponential parameterization of a Hartree--Fock or Kohn--Sham determinant that is based on a generator of single-electron excitations but yet encompasses multi-electron excited determinants. As mentioned in the Introduction, a direct consequence of the use of a nonlinear parameterization is that the properties derived from a response framework will differ from those based on the state-specific formulation. (\textit{ii}) The lack of a $\hat{\Lambda}_{\subm{AB}}(t)$ operator in the exponent of the first equality in Eq.~\eqref{eq:LRCS_parameterization} and thus a direct coupling between direct-product states of the type $\langle i_\subm{A}0_\subm{B}|$ and $|i_\subm{A}j_\subm{B}\rangle$ as well as $\langle i_\subm{A}0_\subm{B}|$ and $|k_\subm{A}j_\subm{B}\rangle $ is the origin of the neglect of state-specific relaxation and London dispersion between the subsystems. (\textit{iii}) The expansion of the time-dependent wave function in the basis of subsystem states means that only intrasubsystem transitions are included, while intersubsystem transitions are excluded. Therefore, a main approximation in the above parameterization is the exclusion of charge-transfer transitions between subsystems in line with our initial restriction on a fixed number of electrons in a given subsystem.  

Having settled on an approximate parameterization of the phase-isolated wave function, we can now construct the
time-dependent quasi-energy of the 	composite system
\begin{align}\label{eq:DP_quasienergy}
Q_{\subm{AB}}(t)= \langle \widetilde{0_{\subm{A}} 0_{\subm{B}}} | \left(\hat{\mathcal{H}} +\hat{V}^t-i\frac{\partial}{\partial t}\right) |\widetilde{0_{\subm{A}} 0_{\subm{B}}}\rangle \ \text{.}
\end{align}
The time-dependent amplitudes are expanded in orders of the perturbation and determined by imposing the variational principle for the time-averaged quasi-energy to be satisfied at the various orders.
Following Eq.~\eqref{eq:TA_quasi_energy}, explicit expressions for response functions of the combined system can then be obtained as perturbation-strength derivatives of the time-averaged quasi-energy in Eq.~\eqref{eq:DP_quasienergy}, evaluated at zero field strengths.
The linear response function becomes
\begin{align}\label{eq:RTCS_LRfunction_compact}
\langle \langle \hat{V}_{\alpha}^{-\omega}; \hat{V}_{\beta}^{\omega} \rangle\rangle =\left.\frac{\text{d}^2\{Q_{\subm{AB}}\}_T}{\text{d}F_{\alpha}^{-\omega}\, \text{d}F_{\beta}^{\omega}}\right|_{\mathbf{F}=\mathbf{0}}= -{\mathbf{V}}_{\alpha}^{\omega\dagger}(\mathbf{E}^{[2]}-\omega \mathbf{S}^{[2]})^{-1}\mathbf{V}_{\beta}^{\omega}\ \text{.}
\end{align}
Ordering the Fourier components of the configuration amplitudes in the operators in Eq.~\eqref{eq:LRCS_parameterization} according to: $\mathbf{\Lambda}^{\omega}=(\bm{\lambda}^{\omega}_{\subm{A}},\, \bm{\lambda}^{\omega\,*}_{\subm{A}},\, \bm{\lambda}^{\omega}_{\subm{B}},\, {\bm{\lambda}^{\omega}_{\subm{B}}}^*)^T$, leads to the following intra- and intersubsystem blocked forms of the vectors and matrices:
\begin{align}
-i\mathbf{V}_{\beta}^{\omega}=\left.\frac{\partial^2\{Q_{\subm{AB}}\}_T}{\partial F_{\beta}^{\omega}\, \partial {\mathbf{\Lambda}^{\omega}}^*}\right|_{\mathbf{F}=\mathbf{0}}=&
-i
\begin{bmatrix}
\mathbf{V}^{\omega}_{\subm{A},\beta}\\
\mathbf{V}^{\omega}_{\subm{B},\beta}
\end{bmatrix}\ \text{,}\label{eq:V}
\end{align}
\begin{align}
\mathbf{E}^{[2]}-\omega\mathbf{S}^{[2]}=&\left.\frac{\partial^2\{Q_{\subm{AB}}\}_T}{\partial {\mathbf{\Lambda}^{\omega}}^*\, \partial {\mathbf{\Lambda}^{\omega}}}\right|_{\mathbf{F}=\mathbf{0}}
=
\begin{bmatrix}
\mathbf{E}_{\subm{A}}^{[2]} & \mathbf{E}_{\subm{AB}}^{[2]}\\
\mathbf{E}_{\subm{BA}}^{[2]} & \mathbf{E}_{\subm{B}}^{[2]}
\end{bmatrix}
-\omega
\begin{bmatrix}
\mathbf{S}_{\subm{A}}^{[2]} & \mathbf{0} \\
\mathbf{0} & \mathbf{S}_{\subm{B}}^{[2]}
\end{bmatrix}.\label{eq:E2}
\end{align}
The diagonal blocks of the electronic Hessian and overlap matrices, arising upon differentiation of the quasi-energy with respect to the wave function parameters belonging to the same subsystem ($\subt{I}=\subt{A},\subt{B}$), take the same overall form as for the isolated subsystems
\begin{align}\label{eq:RTCS_subblock}
\mathbf{E}_{\subm{I}}^{[2]} &= 
\begin{bmatrix}
\mathbf{A}^{\subm{I}} & \mathbf{B}^{\subm{I}} \\
{\mathbf{B}^{\subm{I}}}^* & {\mathbf{A}^{\subm{I}}}^* 
\end{bmatrix}; \hspace{8pt}
\mathbf{S}_{\subm{I}}^{[2]} = 
\begin{bmatrix}
\mathbf{1} & \mathbf{0} \\
\mathbf{0} & -\mathbf{1}
\end{bmatrix};\hspace{8pt}
\mathbf{V}_{\subm{I},\alpha}^{\omega}=
\begin{bmatrix}
\mathbf{g}_{\alpha}^{\subm{I}}\\
-{\mathbf{g}_{\alpha}^{\subm{I}}}^*
\end{bmatrix}\ ,
\end{align}
where the symbols $\mathbf{0}$ and $\mathbf{1}$ are used to denote appropriately sized null and identity matrices. As exemplified for subsystem \subt{A}, the elements of the subsystem blocks take the form
\begin{align}\label{eq:matrices}
A^{\subm{A}}_{ij}=\ &
\langle 0_{\subm{A}} 
0_{\subm{B}} |\left[\hat{q}_{i_{\subm{A}}},\left[\hat{\mathcal{H}}_{\subm{A}}+\hat{\mathcal{V}}_{\subm{AB}},\hat{q}_{j_{\subm{A}}}^{\dagger}\right]\right] |0_{\subm{A}}0_{\subm{B}} \rangle\\
 =\ & \langle i_{\subm{A}} 0_{\subm{B}} | \hat{\mathcal{H}}_{\subm{A}}+\hat{\mathcal{V}}_{\subm{AB}}| j_{\subm{A}} 0_{\subm{B}}\rangle-\delta_{i_{\subm{A}}j_{\subm{A}}}\langle 0_{\subm{A}}0_{\subm{B}} | \hat{\mathcal{H}}_{\subm{A}}+\hat{\mathcal{V}}_{\subm{AB}} | 0_{\subm{A}}0_{\subm{B}}\rangle \notag \\
 =\ & \langle i_{\subm{A}} | \hat{\mathcal{H}}_{\subm{A}}+ \langle 0_{\subm{B}} | \hat{\mathcal{V}}_{\subm{AB}}| 0_{\subm{B}}\rangle | j_{\subm{A}} \rangle-\delta_{i_{\subm{A}}j_{\subm{A}}}\langle 0_{\subm{A}} | \hat{\mathcal{H}}_{\subm{A}}+ \langle 0_{\subm{B}} |\hat{\mathcal{V}}_{\subm{AB}} | 0_{\subm{B}} \rangle | 0_{\subm{A}}\rangle 
\notag\\[0.1in]
B^{\subm{A}}_{ij}=\ &\langle 0_{\subm{A}} 0_{\subm{B}} |\left[\hat{q}_{i_{\subm{A}}},\left[\hat{\mathcal{H}}_{\subm{A}}+\hat{\mathcal{V}}_{\subm{AB}},\hat{q}_{j_{\subm{A}}}\right]\right] |0_{\subm{A}}0_{\subm{B}} \rangle=0 \ , \notag\\[0.1in]
\mathrm{g}_{\alpha,i}^\subm{A}=\ &\langle0_{\subm{A}}0_{\subm{B}}| [\hat{q}_{i_{\subm{A}}},\hat{V}_{\subm{A},\alpha}^{\omega}] |0_{\subm{A}}0_{\subm{B}}\rangle
 =\  \langle0_{\subm{A}}| [\hat{q}_{i_{\subm{A}}},\hat{V}_{\subm{A},\alpha}^{\omega}] |0_{\subm{A}}\rangle \ .
\end{align}
In addition to the isolated subsystem term, the electronic subsystem Hessian in Eq.~\eqref{eq:RTCS_subblock} incorporates a contribution from the intersubsystem coupling. In particular, it describes the coupling between intrasubsystem excitations under the influence of the electrostatic potential produced by the electronic ground state of the other subsystem.
Note that the vanishing $\mathbf{B}^{\subm{I}}$ blocks result from our choice of a CI parameterization within the subsystems. However, we choose to keep these blocks in our above illustration of the structure of the electronic Hessian of the subsystems as they could be nonzero for other wave function models. The generally rectangular off-diagonal block $\mathbf{E}_{\subm{AB}}^{[2]}$ and its conjugate transpose ${\mathbf{E}_{\subm{BA}}^{[2]}}$ in Eq.~\eqref{eq:E2} describe the intersubsystem coupling.
Using the same ordering as in Eq.~\eqref{eq:RTCS_subblock}, the structure of this block can be written as
\begin{align}\label{eq:RTCS_offdiag_subblock}
\mathbf{E}_{\subm{AB}}^{[2]}
=
\begin{bmatrix}
\mathbf{\Gamma} & \mathbf{\Theta} \\
\mathbf{\Theta}^* & \mathbf{\Gamma}^*
\end{bmatrix}\ ,
\end{align}
with elements given by
\begin{align}\label{eq:E2_offdiagonal}
\Gamma_{ij}=\ &\langle 0_{\subm{A}} 0_{\subm{B}} |\left[\hat{q}_{i_{\subm{A}}},\left[\hat{\mathcal{V}}_{\subm{AB}},\hat{q}_{j_{\subm{B}}}^{\dagger}\right]\right] |0_{\subm{A}}0_{\subm{B}} \rangle 
=\ \langle i_{\subm{A}} 0_{\subm{B}} |\hat{\mathcal{V}}_{\subm{AB}}| 0_{\subm{A}} j_{\subm{B}}\rangle\ \ \text{,}\\[0.1in]
\Theta_{ij}=\ &\langle 0_\subm{A} 0_\subm{B} |\left[\hat{q}_{i_\subm{A}},\left[\hat{\mathcal{V}}_\subm{AB},\hat{q}_{j_\subm{B}}\right]\right] |0_\subm{A}0_\subm{B} \rangle
=\ -\langle i_{\subm{A}} j_{\subm{B}} |\hat{\mathcal{V}}_{\subm{AB}}| 0_{\subm{A}} 0_{\subm{B}}\rangle\ .
\end{align}
The off-diagonal blocks couple excitations in one subsystem with those in the other, and as follows from these expressions, the coupling is described through a Coulombic interaction of transition densities in the two subsystems. Note that the off-diagonal blocks $\mathbf{S}_{\subm{AB}}^{[2]}$ and $\mathbf{S}_{\subm{BA}}^{[2]}$ vanish due to the zero-overlap assumption. 

By analogy to exact-state theory, a pole and residue analysis of the linear response function in Eq.~\eqref{eq:RTCS_LRfunction_compact} determines expressions for excitation energies 
and transition strengths for ground- to excited-state
transitions in the combined system for the specific choice of parameterization in Eq.~\eqref{eq:LRCS_parameterization}. Accordingly, excitation energies are found as eigenvalues of the generalized eigenvalue equation involving the electronic Hessian and the metric
in terms of the overlap matrix:\cite{olsen1985linear}
\begin{align}\label{eq:RTCS_eigenvalue_equation}
\mathbf{E}^{[2]}\mathbf{X}&=\mathbf{S}^{[2]}\mathbf{X}\bm{\Omega}\ \text{.}
\end{align}
We recall the well-known feature of this generalized eigenvalue problem where the eigenvalues come in pairs $\pm\omega_n$ and the associated eigenvectors are related,\cite{olsen1985linear,list2014identifying} and we shall use positive and negative indices ($\omega_{-n}=-\omega_n$) to denote paired solutions. Here, $\mathbf{X}$ is the matrix of eigenvectors satisfying the orthonormality relation
\begin{align}\label{eq:orthogonality_full}
 \mathbf{X}^{\dagger}\mathbf{S}^{[2]}\mathbf{X} &=\bm{\sigma};\hspace{20pt} \sigma_{nm}=\text{sgn}(n)\delta_{nm}\ \text{,}
\end{align} 
and $\bm{\Omega}$ is the diagonal matrix containing the associated eigenvalues $\pm \omega_{n}$.

For later reference, we introduce the partitioned form of the eigenvector matrix according to the blocked structure in Eq.~\eqref{eq:E2}
\begin{align}\label{eq:eigenvector}
\mathbf{X}=
\begin{bmatrix}
\mathbf{X}^{\subm{A}} & \mathbf{X}^{\subm{AB}} \\
\mathbf{X}^{\subm{BA}} & \mathbf{X}^{\subm{B}}
\end{bmatrix},
\end{align}
where the off-diagonal blocks describe the degree of delocalization of the excitations in the combined system. The transition strength associated with excitation $n$ in the composite system can be obtained from the residue of the linear response function as
\begin{align}\label{eq:res}
T^{0n}_{\alpha\beta}=\lim_{\omega\rightarrow\omega_{n}}(\omega-\omega_{n})\langle\langle\hat{V}_{\alpha}^{-\omega};\hat{V}_{\beta}^{\omega}\rangle\rangle =
 {\mathbf{V}}_{\alpha}^{-\omega_n}\mathbf{X}_n^{}\mathbf{X}_n^{\dagger}{\mathbf{V}}_{\beta}^{\omega_{n}}\ \text{,}
\end{align}
and used to compute the dimensionless oscillator strength
\begin{align}
f_{0n}=\frac{2\omega_{n}}{3}T_{\alpha\alpha}^{0n}\ \text{.}
\end{align}

Having outlined the response formalism for the combined system, the main objective in the two following subsections is to obtain expressions for subsystem contributions to properties of the combined system by solving effective equations within the subsystem spaces. As will be shown in Section \ref{sec:PE}, this alternative approach to the solution of the full system provides a direct link to polarizable embedding.

\subsubsection{Subsystem decomposition: electronic response properties}
To decompose the linear response function of the combined system into subsystem contributions,
we use that the inverse of a blocked matrix with nonsingular square diagonal blocks ($\mathbf{U}$ and $\mathbf{Z}$) may be written as 
\begin{align}\label{eq:matrixlemma}
\left[\begin{smallmatrix}
\mathbf{U} & \mathbf{V}\\
\mathbf{W} & \mathbf{Z}
\end{smallmatrix}\right]^{-1}=
\left[
\begin{smallmatrix}
\left(\mathbf{U}-\mathbf{V}\mathbf{Z}^{-1}\mathbf{W}\right)^{-1} & -\left(\mathbf{U}-\mathbf{V}\mathbf{Z}^{-1}\mathbf{W}\right)^{-1}\mathbf{V}\mathbf{Z}^{-1}\\
-\left(\mathbf{Z}-\mathbf{W}\mathbf{U}^{-1}\mathbf{V}\right)^{-1}\mathbf{W}\mathbf{U}^{-1} & \left(\mathbf{Z}-\mathbf{W}\mathbf{U}^{-1}\mathbf{V}\right)^{-1}
\end{smallmatrix}\right]\ \text{,}
\end{align}
which corresponds to using the L\"{o}wdin partitioning technique.\cite{lowdin1962bstudies,lowdin1963studies,lowdin1962astudies}
Using this identity, we can rewrite the matrix resolvent in 
Eq.~\eqref{eq:RTCS_LRfunction_compact}
in a partitioned form and obtain the following subsystem decomposition 
of the linear response function:
\begin{align}\label{eq:molprops}
\langle \langle \hat{V}_{\alpha}^{-\omega};\hat{V}_{\beta}^{\omega}\rangle\rangle
=&
- \left(\mathbf{V}^{\omega\dagger}_{\subm{A},\alpha}\widetilde{\mathbf{N}}^{\omega}_{\subm{A},\beta}
+\mathbf{V}^{\omega\dagger}_{\subm{B},\alpha}\widetilde{\mathbf{N}}_{\subm{B},\beta}^{\omega}\right)
\notag\\
=&-\mathbf{V}_{\subm{A},\alpha}^{\omega\dagger}
\left(\widetilde{\mathbf{E}}_{\subm{A}}^{[2]}(\omega)-\omega\mathbf{S}^{[2]}_{\subm{A}}\right)^{-1}\widetilde{\mathbf{V}}_{\subm{A},\beta}^{\omega}-\mathbf{V}_{\subm{B},\alpha}^{\omega\dagger}
\left(\widetilde{\mathbf{E}}_{\subm{B}}^{[2]}(\omega)-\omega\mathbf{S}_{\subm{B}}^{[2]}\right)^{-1}\widetilde{\mathbf{V}}_{\subm{B},\beta}^{\omega}
\ \text{.}
\end{align}
Since the subsystems are treated on the same footing in this representation,  Eq.~\eqref{eq:molprops} will in the following be referred to as the symmetric subsystem decomposition (SD).
In contrast to Eq.~\eqref{eq:RTCS_LRfunction_compact}, the dimensions in the above expression have been reduced to those of the excitation manifolds of the individual subsystems (e.g., $\text{dim}_\subm{A}{\times}\text{dim}_\subm{A} $) instead of that of the full system $(\text{dim}_\subm{A}{+}\text{dim}_\subm{B}){\times}(\text{dim}_\subm{A}{+}\text{dim}_\subm{B})$.
As indicated by tildes, the response vectors $\widetilde{\mathbf{N}}^{\omega}$ in Eq.~\eqref{eq:molprops} are modified quantities that satisfy the following set of effective linear response equations
\begin{align}
\begin{split}
\left(\widetilde{\mathbf{E}}_{\subm{A}}^{[2]}(\omega)-\omega\mathbf{S}_{\subm{A}}^{[2]}\right)\widetilde{\mathbf{N}}_{\subm{A},\beta}^{\omega}&= \widetilde{\mathbf{V}}^{\omega}_{\subm{A},\beta}\ \text{,}\\
\left(\widetilde{\mathbf{E}}_{\subm{B}}^{[2]}(\omega)-\omega\mathbf{S}_{\subm{B}}^{[2]}\right)\widetilde{\mathbf{N}}_{\subm{B},\beta}^{\omega}&= \widetilde{\mathbf{V}}^{\omega}_{\subm{B},\beta}\ \text{.}\label{eq:eff_response_eq}
\end{split}
\end{align}
In addition to the implicit modifications through the polarization of the reference state vectors (changes induced by interactions within the subsystem through the interaction part of the diagonal blocks, e.g.,~$\mathbf{E}_{\subm{A}}^{[2]}$), the presence of the other subsystem manifests in explicit contributions to the electronic Hessian and property gradients. In Eqs.~\eqref{eq:molprops} and \eqref{eq:eff_response_eq}, this has been written compactly by introducing effective Hessians and property gradients defined, here for subsystem \subt{A}, as
\begin{align}
\widetilde{\mathbf{E}}_{\subm{A}}^{[2]}(\omega)&= \mathbf{E}^{[2]}_{\subm{A}}-\mathbf{E}_{\subm{AB}}^{[2]}\left(\mathbf{E}_{\subm{B}}^{[2]}-\omega\mathbf{S}^{[2]}_{\subm{B}}\right)^{-1}\mathbf{E}_{\subm{BA}}^{[2]}\label{eq:effective_electronicHessian}\ \text{,}\\
\widetilde{\mathbf{V}}_{{\subm{A}},\beta}^{\omega} &= \mathbf{V}_{\subm{A},\beta}^{\omega}-\mathbf{E}_{\subm{AB}}^{[2]}\left(\mathbf{E}_{\subm{B}}^{[2]}-\omega\mathbf{S}^{[2]}_{\subm{B}}\right)^{-1}\mathbf{V}_{\subm{B},\beta}^{\omega}\ \text{.}\label{eq:effectivegradients}
\end{align}
These two expressions are key of the present work, since they show how the properties of subsystem \subt{A} are affected by the presence of another interacting subsystem. As follows, the coupling interactions between subsystems are governed by three different kinds of mechanisms. The response vector for subsystem \subt{A} describes changes induced by: (\textit{i}) the indirect coupling through the modification of the pure intrasubsystem term $\mathbf{E}^{[2]}_\subm{A}$ to include the ground-state electrostatic potential of subsystem \subt{B}. (\textit{ii})
The explicit coupling to subsystem \subt{B} through the frequency-dependent term in the electronic Hessian. A similar result was obtained in the work by Neugebauer in which a subsystem partitioning of the eigenvalue problem within the linear response generalization of subsystem DFT is discussed.\cite{neugebauer2007couplings} (\textit{iii}) The direct interaction between the applied external field and subsystem \subt{B} through the modified property gradient.  
In particular, 
we recognize the matrix resolvent of the second term in Eqs.~\eqref{eq:effective_electronicHessian} and \eqref{eq:effectivegradients} as a matrix representation of the frequency-dependent linear polarizability 
of the polarized reference state of subsystem \subt{B} (i.e., self-consistently polarized
by subsystem \subt{A}), evaluated at the optical frequency $\omega$.\cite{hsu2001excitation} This can be seen by rewriting the intersubsystem coupling blocks according to
\begin{align}
\Gamma_{ij}=&\sum_{pq\in \subm{A}}\sum_{rs\in \subm{B}}v_{pq,rs}\langle 0_\subm{A}0_\subm{B} |[\hat{q}_{i_\subm{A}},[\hat{E}_{pq}\hat{E}_{rs},\hat{q}_{j_\subm{B}}^{\dagger}]]|0_\subm{A}0_\subm{B}\rangle\notag\\
=&\sum_{pq\in \subm{A}}\sum_{rs\in \subm{B}}v_{pq,rs}\langle 0_\subm{A}|[\hat{q}_{i_\subm{A}},\hat{E}_{pq}]| 0_\subm{A}\rangle \langle 0_\subm{B} | [\hat{E}_{rs},\hat{q}_{j_\subm{B}}^{\dagger}]|0_\subm{B}\rangle\notag\\
=&-\sum_{pq\in \subm{A}}\sum_{rs\in \subm{B}}v_{pq,rs}\langle 0_\subm{A}|[\hat{q}_{i_\subm{A}},\hat{E}_{pq}]| 0_\subm{A}\rangle \langle 0_\subm{B} | [\hat{q}_{j_\subm{B}}^{\dagger},\hat{E}_{rs}]|0_\subm{B}\rangle\notag\\
=& \text{g}_{\hat{\mathcal{V}}_\mathbf{r},i}^\subm{A}\text{g}_{\hat{\mathcal{\rho}}_\mathbf{r},j}^{\subm{B}*} \ \text{,}
\end{align}
and similarly
\begin{align}
\Theta_{ij}=&-\sum_{pq\in \subm{A}}\sum_{rs\in \subm{B}}v_{pq,rs}\langle 0_\subm{A}|[\hat{q}_{i_\subm{A}},\hat{E}_{pq}]| 0_\subm{A}\rangle \langle 0_\subm{B} | [\hat{q}_{j_\subm{B}},\hat{E}_{rs}]|0_\subm{B}\rangle\notag\\
=&-\text{g}_{\hat{\mathcal{V}}_\mathbf{r},i}^\subm{A}\text{g}_{\hat{\mathcal{\rho}}_\mathbf{r},j}^{\subm{B}}\ \text{,}
\end{align}
where the vectors are defined in the same way as the property gradient in Eq.~\eqref{eq:RTCS_subblock} but with the perturbation operator being replaced by either the electrostatic potential operator or the density operator.
The off-diagonal block of the electronic Hessian can then be written as
\begin{align}\label{eq:rewriteEab}
\mathbf{E}_{\subm{AB}}^{[2]}= \mathbf{V}_{\hat{\mathcal{V}}_\mathbf{r}^{\subm{A}}} \mathbf{V}_{\hat{\rho}_\mathbf{r}^{\subm{B}}}^\dagger =\mathbf{V}_{\hat{{\rho}}_\mathbf{r}^{\subm{A}}} \mathbf{V}_{\hat{\mathcal{V}}_\mathbf{r}^{\subm{B}}}^\dagger \ \text{.}
\end{align}
Substituting this expression into the last term of Eq.~\eqref{eq:effective_electronicHessian}, we obtain
\begin{align}\label{eq:effE2A_offdiag}
\mathbf{E}_{\subm{AB}}^{[2]}\left(\mathbf{E}_{\subm{B}}^{[2]}-\omega\mathbf{S}_{\subm{B}}^{[2]}\right)^{-1}\mathbf{E}_{\subm{BA}}^{[2]}
=& \mathbf{V}_{\hat{\mathcal{V}}_\mathbf{r}^\subm{A}}\biggl[ \mathbf{V}^{\dagger}_{\hat{\rho}_\mathbf{r}^\subm{B}}\left(\mathbf{E}_{\subm{B}}^{[2]}-\omega\mathbf{S}_{\subm{B}}^{[2]}\right)^{-1}\mathbf{V}_{\hat{\rho}^\subm{B}_{\mathbf{r}^{\prime}}}\biggl]\mathbf{V}_{\hat{\mathcal{V}}_{\mathbf{r}^{\prime}}^{\subm{A}}}^\dagger \notag\\
=& \mathbf{V}_{\hat{\mathcal{V}}_\mathbf{r}}^\subm{A} C_{\mathbf{r},\mathbf{r}'}^{\subm{B}}(\omega)    \mathbf{V}_{\hat{\mathcal{V}}_{\mathbf{r}^{\prime}}}   ^{\subm{A}\dagger}\ ,
\end{align}
where the factor in square brackets is identified with the frequency-dependent generalized linear polarizability of the polarized ground state of subsystem \subt{B}:  $C_{\mathbf{r},\mathbf{r}'}^{\subm{B}}(\omega)=\langle\langle \hat{\rho}_\subm{B}(\mathbf{r});\hat{\rho}_\subm{B}(\mathbf{r}^{\prime})\rangle\rangle_{\omega}$, which, upon assuming the dipole approximation for subsystem \subt{B}, reduces to the electric dipole--dipole polarizability tensor $\bm{\alpha}^{\subm{B}}(\omega)$.

The property gradients for subsystem \subt{A} describe the electrostatic potential generated by a transition density. The last term in Eq.~\eqref{eq:effective_electronicHessian} can thus be interpreted as the linear response of subsystem \subt{B} induced by the electrostatic potential due to a transition density of subsystem \subt{A}, which in turn produces an electrostatic potential acting on \subt{A}.
In the same way, the last term of the effective property gradient in Eq.~\eqref{eq:effectivegradients} can be interpreted as the electrostatic potential acting on subsystem \subt{A} due to the linear polarization induced in subsystem \subt{B} by its direct interactions with the external field. In other words, subsystem \subt{B} acts as a source of a field that gives rise to an effective field strength at the location (but devoid) of subsystem \subt{A} different from the external field, as represented by $\widetilde{\mathbf{V}}_{\subm{A},\beta}^{\omega}$ and ${\mathbf{V}}_{\subm{A},\beta}^{\omega}$, respectively. Note that because of the tensorial nature of the environment polarizability, the external field can be screened differently in different directions.
It is also worth underlining the relative distance dependency of these two interaction mechanisms. Assuming a dipole approximation for the interaction operator, it follows that the former effect decays quickly with increasing intermolecular distance $r$, displaying a $r^{-6}$ behavior, whereas the latter effect is comparably long range, dropping of as $r^{-3}$. 
This difference in interaction range is important to consider in practical calculations in order to get a balanced description of the various contributing terms when dealing with finite-sized systems.\cite{rinaldi2013discrete} 

As we shall see later, the second terms of Eqs.~\eqref{eq:effective_electronicHessian} and \eqref{eq:effectivegradients} reduce exactly to the form of the so-called dynamic reaction field and the effective external field (EEF) effects, the latter also referred to as the local field effect, appearing in polarizable embedding.\cite{jensen2005microscopic,list2016excited} 
The effective external field effect is of the same origin as the cavity-field effect in polarizable continuum models leading to effective properties,\cite{botcher1973theory,wortmann1998effective,cammi1998calculation,pipolo2014cavity2} but is defined with respect to the external probing field rather than the Maxwell field within the dielectric.\cite{list2016local}
It should also be noted that the same terms describe the so-called image- and near-field effects in the context of treating molecules adsorbed to metal nanoparticles.\cite{morton2011discrete,payton2012discrete}

Finally, we note that neither of the subsystem contributions to the linear response function 
in Eq.~\eqref{eq:molprops} are symmetric with respect to the left and right property gradients. For that reason, there is no guarantee that the effective subsystem polarizability tensors are symmetric, as discussed before,\cite{neugebauer2009calculation} or that the diagonal elements are positive in the static limit. This is in contrast to the polarizability tensor of the combined system, which is symmetric, as it should be.  Physically, the nonsymmetric form of the individual terms can be viewed as describing the linear response of a property associated with the operator $\hat{V}_{\subm{A},\alpha}^{-\omega}$ to the actual perturbing field acting on subsystem \subt{A} in the presence of subsystem \subt{B}.

\subsubsection{Subsystem decomposition: electronic transition properties}
In this section, we perform an analysis of the transition properties of the composite system as relevant for resonant external fields. As pointed out in the Introduction, the lowest electronic transitions are often localized in nature and can often be attributed predominantly to a particular subsystem, {i.e.}, the excitation vectors have the dominant contribution in the excitation manifold of one of the subsystems.
The condition under which such excitations occur follows from perturbation theory: the magnitude of the coupling strengths of the excited state of interest with electronic states in the other subsystem should be small compared with their energetic separation. This ratio dictates the magnitude difference between the diagonal and off-diagonal blocks in Eq.~\eqref{eq:eigenvector} and thus the extent of the delocalization of a given electronic excitation in the combined system. 

Rather than solving the full generalized eigenvalue equation in Eq.~\eqref{eq:RTCS_eigenvalue_equation}, the excitation energies of the combined system may be determined from the effective response equations in Eq.~\eqref{eq:eff_response_eq} upon zeroing the right-hand sides. Folding the effects of subsystem \subt{B} into the equation for \subt{A} yields the following pseudo generalized eigenvalue equation
\begin{align}\label{eq:eff_eigenvalueeq}
\widetilde{\mathbf{E}}_{\subm{A}}^{[2]}(\omega_n)\widetilde{\mathbf{X}}_n=\omega_n\mathbf{S}_{\subm{A}}^{[2]}\widetilde{\mathbf{X}}_n\ ,
\end{align}
which has the dimension of the excitation manifold of subsystem \subt{A}.
Provided there are no degenerate states located in the \subt{B} part, the excitation energies derived from the reduced system in Eq.~\eqref{eq:eff_eigenvalueeq} are identical to those associated with the subsystem \subt{A} part of the parent system in Eq.~\eqref{eq:RTCS_eigenvalue_equation}, {i.e.}, the full system. This reformulation thus turns the generalized eigenvalue problem for the full system into a dressed subsystem problem, as has been shown before in the subsystem TDDFT framework.\cite{neugebauer2007couplings} In other words, this form allows us to compute the poles associated with the \subt{A} (\subt{B}) block of Eq.~\eqref{eq:RTCS_eigenvalue_equation} without having to consider the full problem with the dimension of both subsystems. 
This is particularly attractive for determining localized transitions as 
relevant from the perspective of embedding calculations. The nonlinearity of Eq.~\eqref{eq:eff_eigenvalueeq} introduced by the frequency-dependent effective Hessian, however, requires knowledge of the solutions beforehand or that
the problem is solved iteratively one excitation at a time, including the construction of a new Hessian for each eigenvalue in question. As a consequence, the subsystem \subt{A} components will not be orthogonal to each other (only the full eigenvectors will).
If only the subsystem term $\mathbf{E}_{\subm{A}}^{[2]}$ is included in Eq.~\eqref{eq:eff_eigenvalueeq}, the result will be denoted by "FP" for frozen ground-state polarization---implying that subsystem \subt{B} is not allowed to respond to the density changes in \subt{A} upon excitation (that is, neglecting intersubsystem couplings).
 As discussed in the previous section, the second term in Eq.~\eqref{eq:effective_electronicHessian} couples an excitation in subsystem \subt{A} with those in \subt{B} by the interaction of the associated transition densities (see Eq.~\eqref{eq:rewriteEab}). 

Let us now return to the evaluation of the transition strengths associated with the excitations in the composite system, but now taking a decomposed form of the linear response function as a starting point. 
As has been shown by Pavanello,\cite{pavanello2013subsystem} each subsystem contribution to the response function (i.e., symmetric decomposition in Eq.~\eqref{eq:molprops}) displays poles at all excitations in the combined system. This complication reflects the delocalization necessarily present in coupled quantum systems, which is not compliant with our heuristic view on local excitations. As a consequence, transition strengths of electronic transitions "localized" in subsystem \subt{A} cannot be identified as the residues of the subsystem \subt{A} contribution to the response function (the first term in Eq.~\eqref{eq:molprops}).

To facilitate the identification of residues, we will explore an alternative partitioning of the matrix resolvent in Eq.~\eqref{eq:RTCS_LRfunction_compact}, which can be obtained by applying the Woodbury matrix identity
(for $\mathbf{U}$ and $\mathbf{Z}$ nonsingular square matrices)
\begin{align}
&(\mathbf{Z}-\mathbf{W}\mathbf{U}^{-1}\mathbf{V})^{-1}\mathbf{W}\mathbf{U}^{-1}=\mathbf{Z}^{-1}\mathbf{W}(\mathbf{U}-\mathbf{V}\mathbf{Z}^{-1}\mathbf{W})^{-1}\ ,\label{eq:woodbury1} \\[0.1in]
&(\mathbf{Z}-\mathbf{W}\mathbf{U}^{-1}\mathbf{V})^{-1}=\mathbf{Z}^{-1}+\mathbf{Z}^{-1}\mathbf{W}(\mathbf{U}-\mathbf{V}\mathbf{Z}^{-1}\mathbf{W})^{-1}\mathbf{V}\mathbf{Z}^{-1}\ , \label{eq:woodbury2}
\end{align}
to the third and fourth blocks of Eq.~\eqref{eq:matrixlemma} (alternatively, the first and second blocks to get the residues related to the \subt{B} part).
Hereby, the linear response function of the combined system can be written as
\begin{align}\label{eq:decomposed}
\langle\langle \hat{V}_{\alpha}^{-\omega};\hat{V}_{\beta}^{\omega}\rangle\rangle=&-\widetilde{\mathbf{V}}^{\omega\dagger}_{\subm{A},\alpha}\left(\widetilde{\mathbf{E}}_{\subm{A}}^{[2]}-\omega\mathbf{S}_{\subm{A}}^{[2] }\right)^{-1}\widetilde{\mathbf{V}}^{\omega}_{\subm{A},\beta}	-\mathbf{V}_{\subm{B},\alpha}^{\omega\dagger}\left({\mathbf{E}}_{\subm{B}}^{[2]}-\omega\mathbf{S}_{\subm{B}}^{[2] }\right)^{-1}\mathbf{V}_{\subm{B},\beta}^{\omega}\ ,
\end{align}
in which the subsystems, in contrast to Eq.~\eqref{eq:molprops}, are treated on an unequal footing. Equation~\eqref{eq:decomposed} will therefore be referred to as the nonsymmetric decomposition (NSD). The same approach has been used in the original formulations of second-order approximate polarization propagator (SOPPA) theory\cite{nielsen1980transition} 
in which the effective quantities describe the doubles correction to the particle-hole spectrum, while the pure double excitations are given by the second term. As will be discussed in Sec.~\ref{sec:PE_response}, the expressions for polarizable embedding in a linear response framework follows, in a similar spirit as SOPPA and algebraic diagrammatic construction,\cite{dreuw2015algebraic} from a perturbation analysis of the individual matrices appearing in Eq.~\eqref{eq:decomposed}.
Note that the second term originates from the first term on the right-hand side of Eq.~\eqref{eq:woodbury2} and corresponds to the linear response function for the ground-state polarized subsystem \subt{B}. An important outcome of this alternative representation of the response function of the combined system is that the poles of the \subt{A}-dominated excitations (all excitations of the composite system) are contained entirely in the first term. However, as a consequence of the appearance of the second term, the partitioning in Eq.~\eqref{eq:decomposed} introduces additional poles in the individual terms at the transition frequencies of the ground-state polarized subsystem \subt{B}. Hence, the correct pole structure for the composite system is only recovered upon taking the sum of the two terms. Nevertheless, the first term in Eq.~\eqref{eq:decomposed} can be used to identify the correct decomposed expression for the residues of the excitations mainly in subsystem \subt{A}, since the second term does not affect excitation energies and transition moments but is needed only in the calculation of the response function.

Until now the normalization of the effective eigenvectors obtained from Eq.~\eqref{eq:eff_eigenvalueeq} has been of no concern since it does not affect excitation energies. However, it is necessary to consider a renormalization of the effective eigenvectors before the transition strengths can be evaluated in the decomposed formulation. In particular, since the eigenvector of the full equation, corresponding to a transition mainly localized in subsystem \subt{A}, is normalized to $\pm 1$ according to
\begin{align}
\mathbf{X}^{\subm{A}\dagger}_n\mathbf{S}_{\subm{A}}^{ [2]}\mathbf{X}^{\subm{A}}_n+
\mathbf{X}^{\subm{BA}\dagger}_n\mathbf{S}_{\subm{B}}^{[2]}\mathbf{X}^{\subm{BA}}_n=\sigma_{n}\ ,
\end{align}
the subsystem \subt{A} component (first term) must have a norm less than unity.
The \subt{A} and \subt{B} components of the eigenvector are related through Eq.~\eqref{eq:RTCS_eigenvalue_equation}, which can be used to rewrite the normalization condition in terms of the \subt{A} component. This leads to the following renormalization factor for the effective eigenvector for the given pole $\omega_n$  
 \begin{align}\label{eq:normalization}
 \Gamma_n^{\subm{A}-1}&=\widetilde{\mathbf{X}}_n^{\subm{A}\dagger}\mathbf{S}_{\subm{A}}^{ [2]}\widetilde{\mathbf{X}}_n^{\subm{A}}+
\widetilde{\mathbf{X}}_n^{\subm{A}\dagger}\mathbf{E}_{\subm{AB}}^{[2]}(\mathbf{E}_{\subm{B}}^{[2]}-\omega_n\mathbf{S}_{\subm{B}}^{[2]})^{-1}\mathbf{S}_{\subm{B}}^{[2]}(\mathbf{E}_{\subm{B}}^{[2]}-\omega_n\mathbf{S}_{\subm{B}}^{[2]})^{-1}\mathbf{E}_{\subm{BA}}^{[2]}\widetilde{\mathbf{X}}_n^{\subm{A}}\ .
\end{align}
The transition strength between the ground state and an excited state mainly located in subsystem \subt{A} can then be written, in the partitioned form, as
\begin{align}\label{eq:symmetric_M}
T_{\alpha\beta}^{0n}=\Gamma_n^{\subm{A}}
\widetilde{\mathbf{V}}_{\subm{A},\alpha}^{-\omega_n}\widetilde{\mathbf{X}}^{\subm{A}}_n
{\widetilde{\mathbf{X}}^{\subm{A}\dagger}_n}
\widetilde{\mathbf{V}}_{\subm{A},\beta}^{\omega_n} \ .
\end{align}
What we have achieved up to this point is to have recast Eqs.~\eqref{eq:RTCS_LRfunction_compact} and \eqref{eq:RTCS_eigenvalue_equation} into dressed subsystem expressions that separate out subsystem contributions to molecular and transition properties of the combined system. As will become clear from the final steps taken in Sec.~\ref{sec:PE}, this provides a justification of the various environmental effects appearing in polarizable embedding. The theoretical analysis further allowed to identify what decomposed form of the response function to use for a specific purpose. We underline that such clarification does not follow from a derivation anticipating the classical description of the environment in the first place, where the Ehrenfest and quasi-energy derivative formulations of response theory lead to different response functions.\cite{list2016local}

Although the subsystem decompositions based on standard response theory are illustrative and provide insight into the mechanisms governing the interaction between subsystems, their practical application for transition properties is limited for several reasons: (\textit{i}) the solution of the pseudo generalized eigenvalue problem in Eq.~\eqref{eq:eff_eigenvalueeq} requires an iterative scheme, (\textit{ii}) a new effective subsystem \subt{A} Hessian has to be constructed for each "eigenvalue" of interest, and (\textit{iii}) the problem is ill-defined if poles in subsystem \subt{B} are too close to the one being solved for. Another important point is that the normalization factor in Eq.~\eqref{eq:normalization}, necessary for the evaluation of absorption intensities, cannot be straightforwardly converted into an effective environment analog as needed when turning to embedding models. 
As will be discussed next, these complications can be avoided by considering the combined system within a complex response theory framework in which absorption properties can be computed without having to resolve the individual excitations, which is also advantageous for systems in which the density of electronic states is high. In addition, such framework clearly illustrates the intensity borrowing that occurs between interacting subsystems.

\subsection{Subsystem decomposition in a complex response framework}\label{sec:damped}
In complex response theory, effects of radiative and nonradiative relaxation mechanisms for the decay of the  excited states are modeled in a phenomenological manner by assigning finite lifetimes ($\tau_n$) to the excited states. This leads to complex-valued response functions that are well-defined across the entire frequency range and thus provides resonant-convergent properties. 
The complex linear response function takes the following form\cite{norman2011perspective, norman2005nonlinear, norman2001near}
\begin{align}
{\langle \langle \hat{V}_{\alpha}^{-\omega};\hat{V}_{\beta}^{\omega}\rangle\rangle}=
-\mathbf{V}^{\omega\dagger}_{\alpha}
(\mathbf{E}^{[2]}-(\omega+i\bm{\gamma})\mathbf{S}^{[2]})^{-1}\mathbf{V}^{\omega}_{\beta}\ \text{.}
\end{align}
In practice, it is customary to adopt a common lifetime, and hence damping parameter $\gamma=(2\tau)^{-1}$, for all excited states such that the so-called relaxation matrix becomes $\bm{\gamma}=\gamma\mathbf{1}$.

Similar to the conventional response framework, the complex linear response function may be expressed in alternative subsystem decomposed forms. Applying Eq.~\eqref{eq:matrixlemma} yields its symmetric subsystem decomposition
\begin{align}
{\langle\langle \hat{V}_{\alpha}^{-\omega};\hat{V}_{\beta}^{\omega}\rangle\rangle}=&-\mathbf{V}_{\subm{A},\alpha}^{\omega\dagger}({\widetilde{\mathbf{E}}}_{\subm{A}}^{[2]}({\omega})-({\omega}+i\gamma)\mathbf{S}_{\subm{A}}^{[2]})^{-1}{\widetilde{\mathbf{V}}}_{\subm{A},\beta}^{\omega}\notag\\
&-\mathbf{V}_{\subm{B},\alpha}^{\omega\dagger}({\widetilde{\mathbf{E}}}_{\subm{B}}^{[2]}({\omega})-({\omega}+i\gamma)\mathbf{S}_{\subm{B}}^{[2]})^{-1}{\widetilde{\mathbf{V}}}_{\subm{B},\beta}^{\omega} \ \text{.}
\end{align}
The effective subsystem vector and matrix quantities are now complex, here given for subsystem \subt{A}:
\begin{align}
{\widetilde{\mathbf{E}}}_\subm{A}^{[2]}({\omega})=&\mathbf{E}_{\subm{A}}^{[2]}-\mathbf{E}_{\subm{AB}}^{[2]}(\mathbf{E}_{\subm{B}}^{[2]}-({\omega}+i\gamma)\mathbf{S}_{\subm{B}}^{[2]})^{-1}\mathbf{E}_{\subm{BA}}^{[2]}\ \text{,}\label{eq:complexHessian}\\
{\widetilde{\mathbf{V}}}^{\omega}_{\subm{A},\alpha}=&\mathbf{V}_{\subm{A},\alpha}^{\omega}-\mathbf{E}_{\subm{AB}}^{[2]}(\mathbf{E}_{\subm{B}}^{[2]}-({\omega}+i\gamma)\mathbf{S}_{\subm{B}}^{[2]})^{-1}\mathbf{V}_{\subm{B},\alpha}^{\omega}\ \text{.}\label{eq:complexPG}
\end{align}
Note that even in the usual case of a real (or purely imaginary) external perturbation, the property gradient will be complex as a result of the damped response of the other subsystem (the second term of Eq.~\eqref{eq:complexPG}).
By further rewriting according to Eqs.~\eqref{eq:woodbury1} and \eqref{eq:woodbury2}, we obtain the nonsymmetric decomposition
\begin{align}
{\langle\langle \hat{V}_{\alpha}^{-\omega};\hat{V}_{\beta}^{\omega}\rangle\rangle} =&-{\widetilde{\mathbf{V}}}_{\subm{A},\alpha}^{\omega\dagger'}({\widetilde{\mathbf{E}}}_{\subm{A}}^{[2]}({\omega})-({\omega}+i\gamma)\mathbf{S}_{\subm{A}}^{[2]})^{-1}{\widetilde{\mathbf{V}}}_{\subm{A},\beta}^{\omega}\notag\\
& -\mathbf{V}_{\subm{B},\alpha}^{\omega\dagger}({\mathbf{E}}_{\subm{B}}^{[2]}-({\omega}+i\gamma)\mathbf{S}_{\subm{B}}^{[2]})^{-1}{\mathbf{V}}_{\subm{B},\beta}^{\omega}\ \text{.} \label{LRF-complex-symmetric}
\end{align}
It should be noted that the conjugate transpose of the effective quantities are here assumed, as indicated by the prime, to act only on vectors and matrices but without changing the sign in front of the damping parameter.

In actual calculations, the value of the response function is determined by solving the complex linear response equation. For the subsystem decompositions, this implies solving the complex analogs of Eq.~\eqref{eq:eff_response_eq}. To be amenable to practical implementation, these may be expressed as 
a coupled set of linear equations for the real and imaginary components\cite{kauczor2011onthe} (indicated by superscripts $R$ and $I$, respectively). By using Eq.~\eqref{eq:effE2A_offdiag} and assuming real wave functions, we obtain the following expression for the effective response equations for subsystem \subt{A}
\begin{widetext}
\begin{align}
\left[\begin{smallmatrix}
{{\mathbf{E}}}_{\subm{A}}^{[2]}-\mathbf{V}_{\hat{\mathcal{V}}^\subm{A}_\mathbf{r}}C_{\mathbf{r},\mathbf{r}'}^{\subm{B},R}(\omega)\mathbf{V}_{\hat{\mathcal{V}}^\subm{A}_{\mathbf{r}'}}^\dagger -\omega\mathbf{S}_{\subm{A}}^{[2]} & \mathbf{V}_{\hat{\mathcal{V}}^\subm{A}_\mathbf{r}}C_{\mathbf{r},\mathbf{r}'}^{\subm{B},I}(\omega)\mathbf{V}_{\hat{\mathcal{V}} ^\subm{A}_{\mathbf{r}'}}^\dagger+\gamma\mathbf{S}_\subm{A}^{[2]} \\[0.1in]
-\mathbf{V}_{\hat{\mathcal{V}}^\subm{A}_\mathbf{r}} C_{\mathbf{r},\mathbf{r}'}^{\subm{B},I}(\omega)\mathbf{V}_{\hat{\mathcal{V}}^\subm{A}
_{\mathbf{r}'}}^\dagger -\gamma\mathbf{S}_\subm{A}^{[2]} &
 {{\mathbf{E}}}_{\subm{A}}^{[2]}-\mathbf{V}_{\hat{\mathcal{V}}^\subm{A}_\mathbf{r}}C_{\mathbf{r},\mathbf{r}'}^{\subm{B},R}(\omega)\mathbf{V}_{\hat{\mathcal{V}}^\subm{A}_{\mathbf{r}'}}^\dagger-\omega\mathbf{S}_{\subm{A}}^{[2]} \\[0.05in]
 \end{smallmatrix}\right]
\left[\begin{smallmatrix}
\widetilde{\mathbf{N}}_{\subm{A},\beta}^{\omega,R}\\[0.1in]
\widetilde{\mathbf{N}}_{\subm{A},\beta}^{\omega,I}
 \end{smallmatrix}\right]
=\left[\begin{smallmatrix}
{\mathbf{V}}_{\subm{A},\beta}^{\omega,R}-\mathbf{V}_{\hat{\mathcal{V}}^\subm{A}_\mathbf{r}}C_{\mathbf{r},\beta}^{\subm{B},R}(\omega) \\[0.1in]
{\mathbf{V}}_{\subm{A},\beta}^{\omega,I}-\mathbf{V}_{\hat{\mathcal{V}}^\subm{A}_\mathbf{r}}C_{\mathbf{r},\beta}^{\subm{B},I}(\omega)
 \end{smallmatrix}\right]\ \text{,}
\end{align}
\end{widetext}
where $C_{\mathbf{r},\beta}^{\subm{B}}(\omega)=\langle \langle \hat{\rho}^{\subm{B}}(\mathbf{r});\hat{V}^{\omega}_{\subm{B},\beta}\rangle\rangle_\omega$. As seen, the coupling between the real and imaginary components of the effective response vector is, in addition to the vacuum contribution from the damping within subsystem \subt{A}, mediated by the imaginary part of the generalized linear polarizability for subsystem \subt{B}.

The real part of the complex electric dipole--dipole polarizability is related to the refractive index of the system and, as follows from energy-loss considerations of the perturbing electromagnetic field, the imaginary part is directly proportional to the linear absorption cross section $\sigma(\omega)$.\cite{boyd2003nonlinear,list2015beyond} For a sample that is isotropic with respect to the light polarization, we have
\begin{align}\label{eq:cross-section}
\sigma(\omega)=\frac{\omega}{3\epsilon_0 c_0}
\text{Im}\left[\rule{0pt}{12pt}\alpha_{\alpha\alpha}(\omega)\right]\ \text{,}
\end{align}
where $\epsilon_0$ is the vacuum permittivity and $c_0$ is the speed of light in vacuum. 

The physical significance of the imaginary component of the polarizability can alternatively be recognized by the relation between the integrated absorption cross section and the sum of oscillator strengths (see Appendix \ref{app:conservationlaw})
\begin{align}\label{eq:conservationlaw}
I=\int_0^{\infty}\sigma(\omega) \ \text{d}\omega=\frac{\pi}{2\epsilon_0 c_0}\sum_{n>0}f_{n0} \ \text{.}
\end{align}
In the framework of exact state theory or variational approximate state theory in the complete basis set limit, the Thomas--Reiche--Kuhn sum rule further implies that the sum of oscillator strengths is equal to the number of the electrons in the system $N_\mathrm{e}$. 
Accordingly, for the combined system, we have
\begin{align}\label{eq:sumrule_AB}
\sum_{n>0}f_{n0}^{\subm{AB}}=N_\mathrm{e}^\subm{A}+N_\mathrm{e}^\subm{B}\ \text{,}
\end{align}
and likewise for the subsystems
\begin{align}\label{eq:sumrule_unperturbed}
\sum_{n>0}f_{n0}^{\text{vac},\subm{I}}=\sum_{n>0}f_{n0}^{\text{FP},\subm{I}}=N_\mathrm{e}^\subm{I}; \hspace{15pt}I=\subt{A},\subt{B} \ .
\end{align}
Together, Eqs.~\eqref{eq:conservationlaw} and \eqref{eq:sumrule_AB} reflect that excitations in one subsystem can borrow intensity from transitions in the other subsystem, while the total integrated cross section is preserved.
By combining Eqs.~\eqref{eq:conservationlaw}--\eqref{eq:sumrule_unperturbed} and recalling that the second term of the complex linear response function in the nonsymmetric subsystem decomposition in Eq.~\eqref{LRF-complex-symmetric} is identical to the linear response functions of subsystem \subt{B} within the FP approximation, it follows that 
\begin{align}\label{eq:conservationlaw2}
\int \sigma^{\text{NSD}}_1(\omega)\ \text{d}\omega=\frac{\pi}{2\epsilon_0 c_0}N_\mathrm{e}^\subm{A}; \hspace{25pt}\int \sigma^{\text{NSD}}_2(\omega)\ \text{d}\omega=\frac{\pi}{2\epsilon_0 c_0}N_\mathrm{e}^\subm{B}\ \text{,}
\end{align}
where $\sigma^{\text{NSD}}_1$ and $\sigma^{\text{NSD}}_2$ denote the contribution to the absorption cross section from the first and second term in Eq.~\eqref{LRF-complex-symmetric}, respectively.
That is, if an \subt{A}-dominated transition gains in intensity due to the coupling to excitations in subsystem \subt{B}, then $\sigma^{\text{NSD}}_1(\omega)$ will take on negative values around poles in subsystem \subt{B}. Consequently, in these regions of the spectrum, $\sigma^{\text{NSD}}_1(\omega)$ in itself cannot be associated with an absorption spectrum but it rather becomes imperative to consider the total absorption cross section $\sigma(\omega)$. This discussion is important due to its implications in the context of polarizable embedding (see next section) in which one focuses on the calculation of $\sigma^{\text{NSD}}_1(\omega)$ and, as we have seen, caution is called for in the interpretation of the results of such a calculation.

\section{Polarizable Embedding}\label{sec:PE}
Having derived the expressions for the direct-product \textit{ansatz} for the combined system, we will in this section detail the additional steps that lead to the definition of the PE model, following the work of \'{A}ngy\'{a}n for the derivation of the effective embedding operator.\cite{angyan1992common} In particular, the subsystems will now be treated at different levels, where a classical description will be adopted for subsystem \subt{B}. To reflect this distinction, subsystem \subt{A} will in this section be referred to as the quantum region and \subt{B} as the environment. 
Furthermore, instead of considering the environment as a whole the individual subsystems constituting the environment ${B}=\{b_1,b_2, \dots,b_{N-1}\}$ will be treated separately by decomposing the environment wave function into a product of subsystem contributions, still assuming nonoverlapping subsystem charge densities. We will assume that the unperturbed (i.e., vacuum) environment subsystem eigenfunctions and -energies $\{|0_{{b}}^{(0)}\rangle,|j_{{b}}^{(0)}\rangle\}$ and $\{E_{0_b}^{(0)},E_{j_b}^{(0)}\}$, for $b\in\subt{B}$, are known, where superscripts $(n)$ specify the order with respect to the perturbation (see below).

\subsection{Working Equations}
In the PE model, we invoke a perturbation treatment of all but subsystem \subt{A} and assume that the environment is only linearly responsive. This corresponds to requiring that Eq.~\eqref{eq:effective_equations} for $b\in \subt{B}$
is fulfilled only through first order in terms of the electrostatic potentials from the ground states of the other subsystems. 
Within this approximation, the interaction operator acting on subsystem \subt{A} takes the form
\begin{align}\label{eq:int_operator_perturbation}
\hat{\mathcal{V}}^{\text{int}}=&\underbrace{\sum_{pq\in\subm{A}}\sum_{b\in\subm{B}}\sum_{m\in b }^{M_b}Z_m\bigl[v_{pq}(\mathbf{R}_m)+\sum_{rs\in {b}}v_{pq,rs}D_{{b},rs}^{\scriptsize(0)}\bigr]\hat{E}_{pq}}_{\hat{\mathcal{V}}^{\text{es}}}+\underbrace{\sum_{pq\in\subm{A}}\sum_{b\in\subm{B}}\sum_{rs\in b}v_{pq,rs}D_{b,rs}^{\scriptsize(1)}\hat{E}_{pq}}_{\hat{\mathcal{V}}^{\text{ind}}}\ \text{,}
\end{align}
where an element of the zeroth- and first-order electronic 
densities of subsystem $b$ are defined as 
$D_{b,rs}^{(0)}=\langle 0_{b}^{(0)}|\hat{E}_{rs}|0_{b}^{(0)}\rangle$ and $D_{b,rs}^{(1)}=\langle 0_{b}^{(1)}|\hat{E}_{rs}|0_{b}^{(0)}\rangle +\langle 0_{b}^{(0)}|\hat{E}_{rs}|0_{b}^{(1)}\rangle$, respectively.
The effective interaction operator acting on subsystem \subt{A} consists of contributions from the permanent and induced charge distributions, $\hat{\mathcal{V}}^{\text{es}}$ and $\hat{\mathcal{V}}^{\text{ind}}$, respectively, of the environment subsystems.
In terms of the first-order reduced density and electrostatic potential operators in Eqs.~\eqref{eq:den_operator} and \eqref{eq:pot_operator}, they read
\begin{align}
\hat{\mathcal{V}}^{\text{es}} &=
\sum_{b\in\subm{B}} \int \hat{\rho}_{\subm{A}}^{\text{e}}(\mathbf{r})\langle\hat{\mathcal{V}}_{b}(\mathbf{r})\rangle_{0_{b}}^{(0)} \,\text{d}\mathbf{r}\ \text{,}\label{eq:operators1}\\
\hat{\mathcal{V}}^{\text{ind}} &=
\sum_{b\in\subm{B}} \int \hat{\rho}_{\subm{A}}^{\text{e}}(\mathbf{r}) \langle \hat{\mathcal{V}}^{b}(\mathbf{r})\rangle_{0_{b}}^{(1)}\,\text{d}\mathbf{r}\ \text{,}\label{eq:operators2}
\end{align}
where superscript "\text{e}" signifies that only the electronic part of the operator is included. We have further introduced a shorthand notation for expectation values, e.g.,~ $\langle\hat{\mathcal{V}}_{b}(\mathbf{r})\rangle_{0_{b}}^{(0)}=\langle 0_b^{(0)} | \hat{\mathcal{V}}_b(\mathbf{r})| 0_b^{(0)}\rangle$.
The $\hat{\mathcal{V}}^{\text{es}}$ operator is straightforwardly constructed from the charge densities of the unperturbed environment subsystems and contains the contributions from both the nuclei and electrons in the environment. The induction operator requires the first-order densities to be known. 
Using standard Rayleigh--Schr\"{o}dinger perturbation theory yields the following first-order 
perturbation expression for the 
wave function of the environment subsystems\cite{angyan1992common}
\begin{align}\label{eq:firstorder_correction}
\forall b\in\subt{B}:\hspace{15pt}|0_{b}^{(1)}\rangle =& -\sum_{j> 0} |j_{b}^{(0)}\rangle\int \frac{\langle j_{b}^{(0)} | \hat{\mathcal{V}}_{b}(\mathbf{r})|0_{b}^{(0)}\rangle}{(E_{j_{b}}^{(0)}-E_{0_{b}}^{(0)})}\notag\\[0.05in]
&\times\biggl(\langle \hat{\rho}_{\subm{A}}(\mathbf{r}) \rangle_{0_{\subm{A}}} + \sum_{b'\in\subt{B}\backslash{b}}\left[\langle \hat{\rho}_{{b'}}(\mathbf{r})\rangle_{0_{b'}}^{(0)} +\langle \hat{\rho}_{{b'}}(\mathbf{r})\rangle_{0_{{b'}}}^{(1)}\right] \biggl)\,\text{d}\mathbf{r}\ \text{,}
\end{align}
where the expectation value involving subsystem \subt{A} is over the fully polarized state within the framework of Eq.~\eqref{eq:effective_equations}.
As pointed out by Stone,\cite{stone1989induction} this expression is however inconsistent with the first-order perturbation analysis:  
through the coupling to the other environment subsystems in the third term, Eq.~\eqref{eq:firstorder_correction} contains
contributions to infinite order in the electrostatic potential generated by subsystem \subt{A}.
A strict first-order expression can be obtained by neglecting the many-body polarization among the 
environment subsystems, {i.e.}, removing the first-order term of Eq.~\eqref{eq:firstorder_correction}.

Substituting the first-order correction to the wave function in Eq.~\eqref{eq:firstorder_correction} into $\hat{\mathcal{V}}^{\text{ind}}$
and using the definition of the generalized static polarizability, $C_{\mathbf{r},\mathbf{r}'}^{b,(0)}(\omega=0)$,
of the 
zeroth-order ground states of the environment subsystems, the first-order induction operator can be written as 
\begin{align}\label{eq:induction_operator}
\hat{\mathcal{V}}^{\text{ind}} =& - \sum_{b\in\subm{B}}\int\hat{\rho}_{\subm{A}}^{\text{e}}(\mathbf{r}) \int \left[\iint \frac{C_{\mathbf{r}'',\mathbf{r}'''}^{b,(0)}(0)}{|\mathbf{r}-\mathbf{r}''||\mathbf{r}'-\mathbf{r}'''|}\,\text{d}\mathbf{r}''\,\text{d}\mathbf{r}'''\,\right]\notag\\
&\times 
 \biggl(\langle \hat{\rho}_{\subm{A}}(\mathbf{r}') \rangle_{0_{\subm{A}}} + \sum_{b'\in\subm{B}\backslash b}\left[\langle \hat{\rho}_{{b'}}(\mathbf{r}')\rangle_{0_{{b'}}}^{(0)} +\langle \hat{\rho}_{{b'}}(\mathbf{r}')\rangle_{0_{{b'}}}^{(1)}\right] \biggl)\text{d}\mathbf{r}\,\text{d}\mathbf{r}'\ \text{.} 
\end{align}

In the PE model, the charge distributions of the environment subsystems are represented by multipole expansions rather than densities.
To this end, it is expedient to use a 3-dimensional multi-index notation in which a multi-index $k=(k_x,k_y,k_z)$ is an ordered list of positive integers.\cite{saint1991elementary,olsen2012thesis} The norm and factorial 
of a multi-index are defined as $|k|=k_x+k_y+k_z$ and $k!=k_x!\cdot k_y!\cdot k_z!$, respectively, the sum of two multi-indexes as $k\pm l=(k_x\pm l_x,k_y\pm l_y,k_z\pm l_z)$, and the multi-index power of a vector as
$\mathbf{r}^k=x^{k_x}\cdot y^{k_y}\cdot z^{k_z}$.
Using this notation, a Taylor series expansion of the electrostatic potential operator can be written as\cite{stone2013theory}
\begin{align}\label{eq:taylor_expansion}
\frac{1}{|\mathbf{R}_j-\mathbf{R}_i|}=
\sum_{|k|=0}^{\infty}\frac{(-1)^{|k|}}{k!}\left(\nabla_j^k \frac{1}{|\mathbf{R}_j-\mathbf{R}_{\text{o}}|} \right)(\mathbf{R}_i-\mathbf{R}_{\text{o}})^k\ ,
\end{align}
where $\mathbf{R}_{\text{o}}$ is the expansion point and the summation over $k$ runs over the $\tfrac{1}{2}(|k|+1)(|k|+2)$ Cartesian components.
The multipole form of the potential operator involves components of the Cartesian interaction tensors defined as derivatives of the potential operator\cite{buckingham1967,stone2013theory} 
\begin{align}\label{eq:T}
T_{ij}^{(k)} = \partial^{k}_j\frac{1}{|\mathbf{r}_j-\mathbf{r}_i|} \ \text{;} \qquad \partial^k_j = \frac{\partial^{|k|}}{\partial x_j^{k_x}\partial y_j^{k_y}\partial z_j^{k_z}} \ \text{,}
\end{align}
where the superscript multi-index notation should not be confused with the perturbation order.
Note that in writing Eq.~\eqref{eq:taylor_expansion}, we have used the symmetry properties of the 
interaction tensors, that is,~$T_{ij}^{(k)}{=}({-}1)^{|k|}T_{ji}^{(k)}$. Applying this to the electrostatic potential operators for the environment subsystems yields
\begin{align}\label{eq:pot_operator_expanded}
\hat{\mathcal{V}}_{b}(\mathbf{R}_{b'})=\sum_{|k|=0}^{\infty}\frac{(-1)^{|k|}}{k!}T_{{b}{b'}}^{(k)}\hat{M}_{{b}}^{(k)}(\mathbf{R}_{b})\ \text{,}
\end{align}
where the expansion point $\mathbf{R}_{b}$ is chosen to reside within the 
charge density of subsystem $b$.
A Cartesian component of the $|k|$'th-order multipole moment operator acting on subsystem $b$
is given by
\begin{align}\label{eq:multipole_moment_operator}
\hat{M}_{b}^{(k)}(\mathbf{R}_{b})&= \int\hat{\rho}_{b}(\mathbf{r})(\mathbf{r}-\mathbf{R}_{b})^k \,\text{d}\mathbf{r}\notag\\
&=\sum_{m\in {b}}^{M_{b}}Z_m(\mathbf{R}_m-\mathbf{R}_{b})^k+\sum_{rs\in {b}}m_{rs}^{(k)}(\mathbf{R}_{b})\hat{E}_{rs}\ \text{,}
\end{align}
where the associated electronic multipole integral is given by
\begin{align}
m_{rs}^{(k)}(\mathbf{R}_{b})=-\int\phi_r^{*}(\mathbf{r})(\mathbf{r}-\mathbf{R}_{b})^k\phi_s(\mathbf{r})\ \text{d}\mathbf{r} \ \text{.}
\end{align}
The expectation value of Eq.~\eqref{eq:multipole_moment_operator} corresponds to an electric multipole moment, $M^{(k)}_{b}(\mathbf{R}_{b})$, of subsystem $b$. For instance, letting $|k|=0$ gives the charge, while $|k|=1$ covers the three Cartesian components, i.e., $(1,0,0)$, $(0,1,0)$ and $(0,0,1)$, of a 
dipole moment. Only the lowest-order nonvanishing
multipole moment of the permanent charge distribution 
is independent of the choice of origin (here $\mathbf{R}_{b}$). For the sake of brevity, 
we will suppress this explicit origin dependence in the following.
Finally, substitution of Eq.~\eqref{eq:pot_operator_expanded}
into $\hat{\mathcal{V}}^{\text{es}}$ yields the multipole-expanded
form of the electrostatic interaction operator
\begin{align}\label{eq:electrostatic_operator_expanded}
\hat{\mathcal{V}}^{\text{es}}=\sum_{pq\in {\subm{A}}}\sum_{{b}\in\subm{B}}\sum_{|k|=0}^{\infty}\frac{(-1)^{|k|}}{k!}M_{b}^{(k)}t_{pq}^{(k)}(\mathbf{R}_{b})\hat{E}_{pq}\ \text{,}
\end{align}
where the electrostatic potential integral is defined over the interaction tensors in Eq.~\eqref{eq:T} as
\begin{align}
t_{pq}^{(k)}(\mathbf{R}_{b})=-\int\phi_p^*(\mathbf{r}_i)T_{{b}i}^{(k)}\phi_q(\mathbf{r}_i)\,\text{d}\mathbf{r}_i \ \text{,}
\end{align}
where the index $i$ refers to an electronic coordinate.
Comparing Eq.~\eqref{eq:electrostatic_operator_expanded} to the nonexpanded form in Eq.~\eqref{eq:effective_equations},
it is clear that the two-electron integrals have been 
replaced by one-electron integrals for subsystem \subt{A} over the 
interaction tensors multiplied by the permanent multipole moments of the 
environment subsystems. 

We proceed in the same way for the induction operator in Eq.~\eqref{eq:induction_operator} by replacing all occurrences of the potential
operator by its Taylor series expansion. We further define a component of the 
$|k|$'th-order induced multipole moment belonging to an environment subsystem $b$
as
\begin{align}\label{eq:induced_multipole_moments}
\bar{M}_{b}^{(k)}(\mathbf{R}_{b})&=\int\langle \hat{\rho}_{b}(\mathbf{r})\rangle_{0_{b}}^{(1)} (\mathbf{r}-\mathbf{R}_{b})^k\, \text{d}\mathbf{r} \notag\\
&= \langle 0_{b}^{(1)}| \hat{M}_{b}^{(k)}|0_{b}^{(0)}\rangle+ \langle 0_{b}^{(0)}| \hat{M}_{b}^{(k)}|0_{b}^{(1)}\rangle\ \text{,}
\end{align}
where we have introduced the bar notation to distinguish induced multipole moments from their permanent counterparts.
By contrast to the permanent moments, there are no induced monopoles 
and the nuclear contributions vanish irrespective of the multipole order. 
This is a result of the intermediate normalization of the 
corrections to the wave functions of the environment subsystems. As detailed in Appendix \ref{app:inductionoperator},
the induction part of the interaction operator can be written in the multipole-expanded
form as
\begin{align}
\hat{\mathcal{V}}^{\text{ind}} =&
-\sum_{b\in\subm{B}}\sum_{|k|=1}^{\infty}\frac{1}{k!}\hat{F}_{\subm{A}}^{\text{e},(k)}(\mathbf{R}_{b})\bar{M}_{b}^{(k)}\label{eq:induction_operator_expanded_2}\\
=&-\sum_{b\in\subm{B}}\sum_{|k|=1}^{\infty}\sum_{|l|=1}^{\infty}\frac{1}{k!\cdot l!}\hat{F}_{\subm{A}}^{\text{e},(k)}(\mathbf{R}_{b})P_{b}^{(k,l)}\notag\\
&\times
\left(\langle \hat{F}_{\subm{A}}^{(l)}(\mathbf{R}_{b})\rangle_{0_{\subm{A}}}+\sum_{{b}'\in\subm{B}\backslash b}\left[\langle \hat{{F}}_{b'}^{(l)}(\mathbf{R}_{b})\rangle^{(0)}_{0_{{b}'}}+\langle \hat{{F}}_{{b}'}^{(l)}(\mathbf{R}_{b})\rangle_{0_{{b}'}}^{(1)}\right] \right)\ \text{,}\label{eq:induction_operator_expanded}
\end{align}
where the two alternative expressions arise by 
expanding Eqs.~\eqref{eq:operators2} and \eqref{eq:induction_operator}, 
respectively. 
$P_{{b}}^{(k,l)}$ are static electronic polarizabilities of subsystem $b$, analogous to that in Eq.~\eqref{eq:effE2A_offdiag}, defined as
\begin{align}\label{eq:responsefunction_expanded}
P_{{b}}^{(k,l)}=&\sum_{j_{b}\neq 0_{b}}\Biggl[\frac{\langle 0_{b}^{(0)} | \hat{M}_{b}^{(k)}| j_{b}^{(0)}\rangle \langle j_{b}^{(0)} | \hat{M}_{b}^{(l)}|0_{b}^{(0)}\rangle }{(E_{j_{b}}^{(0)}-E_{0_{b}}^{(0)})}+\frac{\langle 0_{b}^{(0)} | \hat{M}_{b}^{(l)}| j_{b}^{(0)}\rangle \langle j_{b}^{(0)} | \hat{M}_{b}^{(k)}|0_{b}^{(0)}\rangle }{(E_{j_{b}}^{(0)}-E_{0_{b}}^{(0)})}\Biggl]\ \text{,}
\end{align}
recalling that this definition employs the traced multipole moment operators.
Furthermore, $\hat{F}_{\subm{A}}^{(k)}(\mathbf{R}_{b})$ is the $(k{-}1)$'th-order electric-field derivative,
which probes the field derivative produced by subsystem \subt{A} at point $\mathbf{R}_{b}$
\begin{align}\label{eq:field_operator}
\hat{F}_{\subm{A}}^{(k)}(\mathbf{R}_{b})=\underbrace{-\sum_{n\in {\subm{A}}}^{M_{\subm{A}}}Z_nT_{n{b}}^{(k)}}_{F_{\subm{A}}^{\text{n},(k)}(\mathbf{R}_{b})}+\underbrace{(-1)^{|k|+1}\sum_{pq\in {\subm{A}}} t_{pq}^{(k)}(\mathbf{R}_{b})\hat{E}_{pq}}_{\hat{F}_{\subm{A}}^{\text{e},(k)}(\mathbf{R}_{b})}\ \text{,}
\end{align}
as customary defining the field ($|k|=1$) as minus the gradient of the electrostatic potential. The operator
has been partitioned into nuclear and electronic contributions. 
For the environment subsystems, we invoke 
a multipole expansion of the field operator by analogy to Eq.~\eqref{eq:pot_operator_expanded}. Accordingly, we can write 
the zeroth- and first-order 
electric-field derivatives of the environment subsystems in Eq.~\eqref{eq:induction_operator_expanded}, 
in terms of static and induced multipole
moments, respectively, as
\begin{align}
\langle \hat{F}_{{b}'}^{(l)}(\mathbf{R}_{b})\rangle_{0_{{b}'}}^{(0)}&=\sum_{|k|=0}^{\infty}\frac{(-1)^{|k|+1}}{k!}T^{(k+l)}_{{b}'{b}}M_{{b}'}^{(k)}\ \text{,}\notag\\
\langle \hat{F}_{{b}'}^{(l)}(\mathbf{R}_{b})\rangle_{0_{{b}'}}^{(1)}&=\sum_{|k|=1}^{\infty}\frac{(-1)^{|k|+1}}{k!}T^{(k+l)}_{{b}'{b}}\bar{M}_{{b}'}^{(k)}\ \text{.}
\end{align}
Note the difference between the lower limits of the two summations.
Finally, by equating the right-hand sides~of Eqs.~\eqref{eq:induction_operator_expanded_2} and \eqref{eq:induction_operator_expanded}, we
obtain the equation determining the induced multipole moments
\begin{align}\label{eq:induced_dipole_equation}
\forall {b}\in\subt{B}: \notag\\ \bar{M}_{b}^{(k)} =& \sum_{|l|=1}^{\infty}\frac{1}{l!}P_{b}^{(k,l)}F^{\text{tot}(l)}(\mathbf{R}_{b})\\
=&\sum_{|l|=1}^{\infty}\frac{1}{l!}P_{b}^{(k,l)}\Biggl(\langle \hat{F}_{\subm{A}}^{(l)}(\mathbf{R}_{b})\rangle_{0_{\subm{A}}}
+\sum_{{b}'\in\subm{B}\backslash {b}}\biggl[\langle \hat{F}_{{b}'}^{(l)}(\mathbf{R}_{b})\rangle_{0_{b'}}^{(0)}+\sum_{|m|=1}^{\infty}\frac{(-1)^{|m|+1}}{m!}T_{{b}'{b}}^{(m+l)}\bar{M}_{{b}'}^{(m)}\biggl]\Biggl)\ \text{,}\notag
\end{align}
where $F^{\text{tot}(l)}(\mathbf{R}_{b})$ is the total $(l{-}1)$'th-order electric-field derivative acting on subsystem $b$.
As follows from the second equality, it consists of the physical electric-field contributions from the nuclei and electrons in subsystem \subt{A} and 
the permanent multipoles of the other environment subsystems, collectively denoted $F^{(l)}(\mathbf{R}_{b})$, as well as the contribution from the remaining first-order induced multipole moments.
Hence, the first term describes the mutual coupling between subsystem \subt{A}
with all environment subsystems, whereas the second and third terms account for the mutual polarization
between the environment subsystems.

In practice, the multipole expansions in Eqs.~\eqref{eq:electrostatic_operator_expanded} and \eqref{eq:induction_operator_expanded_2} are terminated at a finite order $K_s$, and to improve the convergence properties of the multipole representation distributed multipole expansions (using $S=\sum_{b\in \subm{B}}S_b$ to denote the total number of expansion points) are used instead of one-center expansions. For the expansion over induced moments in Eqs.~\eqref{eq:induction_operator_expanded_2} and \eqref{eq:induced_dipole_equation}, the dipole approximation is introduced and only the dipole--dipole polarizability tensor is taken into account. In this case, Eq.~\eqref{eq:induced_dipole_equation} gives a set of coupled equations determining the induced dipole moments\cite{applequist1972atom}
\begin{align}\label{eq:muind}
\bar{\bm{\mu}}_s(0)= \sum_{s'=1}^S\bm{\mathcal{R}}_{ss'}(0)\mathbf{F}(\mathbf{R}_{s'})\ \text{,} 
\end{align}
where the polarizability tensors for the individual sites have been 
replaced by a $(3S{\times}3S)$-dimensional classical linear response matrix (also known as the relay matrix) given by
\begin{align}
{\bm{\mathcal{R}}}(\omega) =
 \left(\begin{array}{cccc} \bm{\alpha}_1(\omega)^{-1} & -\mathbf{T}^{(2)}_{12}&\cdots  & -\mathbf{T}^{(2)}_{1S} \\[1.0ex]
                                   -\mathbf{T}^{(2)}_{21} & \bm{\alpha}_{2}(\omega)^{-1} & \ddots & \vdots            \\[1.0ex]
                                    \vdots & \ddots & \ddots & -\mathbf{T}^{(2)}_{(S-1)S}\\[1.0ex]
                                    -\mathbf{T}^{(2)}_{S1}    & \cdots   &-\mathbf{T}^{(2)}_{S(S-1)} &  \bm{\alpha}_{S}(\omega)^{-1}         \\[1.0ex]
\end{array}  \right)^{-1}.\label{eq:relaymatrix}
\end{align}  
This matrix holds the inverse of the distributed 
electronic dipole--dipole polarizability tensors on the diagonal
and second-order interaction tensors in the off-diagonal blocks. 
Upon contraction with unit vectors, $\bm{\mathcal{R}}(\omega)$ models the dipole--dipole polarizability of the environment.

By combining Eqs.~\eqref{eq:electrostatic_operator_expanded} and \eqref{eq:induction_operator_expanded} in truncated and distributed forms, we finally obtain the embedding operator defining the PE model:
\begin{align}\label{eq:PE_operator}
\hat{v}_{\text{PE}}=&\sum_{pq\in {\subm{A}}}\sum_{s=1}^S\sum_{|k|=0}^{K_s}\frac{(-1)^{|k|}}{k!}M_{s}^{(k)}t_{pq}^{(k)}(\mathbf{R}_{s})\hat{E}_{pq}-\sum_{s=1}^S\bar{{\mu}}_{s,\alpha}(0)\hat{{F}}^{\text{e}}_{\subm{A},\alpha}(\mathbf{R}_s)\ \text{.}
\end{align} 
The induced dipoles, and in turn the embedding operator, depend on the wave function of subsystem \subt{A} through the electric fields. In other words, upon averaging over the environment wave functions, the Hamiltonian turns into a nonlinear effective operator.

\subsection{Response theory framework}\label{sec:PE_response}
The extension of the PE model within a quantum-mechanical response framework usually proceeds by assuming the classical description of the environment from the outset, i.e.,~starting from Eq.~\eqref{eq:PE_operator} and the associated energy functional (see, e.g., Ref.~\citenum{list2016local}). For the present analysis, we will instead begin from the response expressions derived in Sec.~\ref{sec:compositesystems} and outline the additional assumptions that lead to the expressions within the PE framework. 
As briefly alluded to in Sec.~\ref{sec:compositesystems}, the differentiated treatment in the PE model is achieved by a perturbation analysis of the quantities in the linear response function, although, as we will presently discuss, the choice of truncation is not fully coherent from a perturbation theory point of view.
Since special attention is given to subsystem \subt{A}, and the sole purpose of subsystem \subt{B} in this context is to obtain a realistic description of the properties of \subt{A}, the order counting will be performed on the effective subsystem \subt{A} quantities.

As a first step toward the PE model, one includes terms in the pure subsystem \subt{A} blocks and vectors through second order (in the meaning of Eq.~\eqref{eq:firstorder_correction}), using a first-order corrected wave function for the environment that is normalized through second order.
The pure \subt{B} blocks as well as the coupling blocks are evaluated only through lowest nonvanishing order. That is, only the zeroth-order contribution to the wave function of the environment subsystem is included. For the electronic Hessian, this implies that the $\mathbf{E}_{\subm{B}}^{[2]}$ matrix must be known through zeroth order and the $\mathbf{E}_\subm{AB}^{[2]}$ matrix through first order. Accordingly, the electronic Hessian and metric matrices are approximated as
\begin{align}
\mathbf{E}^{[2]}=
\begin{bmatrix}
\mathbf{E}_\subm{A}^{[2]\text{(0,1,2)}} & \mathbf{E}_\subm{AB}^{[2]\text{(1)}}\\
\mathbf{E}_\subm{AB}^{[2]\text{(1)}} & \mathbf{E}_\subm{B}^{[2]\text{(0)}}
\end{bmatrix}; \hspace{10pt}
\mathbf{S}^{[2]}=
\begin{bmatrix}
\mathbf{S}_{\subm{A}}^{[2](0,1,2) } & \mathbf{0}\\
\mathbf{0} & \mathbf{S}_{\subm{B}}^{[2](0)}
\end{bmatrix}.
\end{align}
Although the pure \subt{B} block is treated only through zeroth order, its effect on excitations in \subt{A} is correct through second order as can be seen from the resulting effective electronic Hessian for subsystem \subt{A} 
\begin{align}\label{eq:electronicHessian_II}
\widetilde{\mathbf{E}}_\subm{A}^{[2]}\stackrel{\,^{\mathrm{(II)}}}{=}&\underbrace{\mathbf{E}_\subm{A}^{[2]\text{(0,1,2)}}}_{\text{I}}-\underbrace{\mathbf{E}_\subm{AB}^{[2]\text{(1)}}(\mathbf{E}_\subm{B}^{[2](0)}-\omega\mathbf{S}_{\subm{B}}^{[2](0)})^{-1}\mathbf{E}_\subm{BA}^{[2]\text{(1)}}}_{\text{II}}\ \text{.}
\end{align}
This order truncation thus provides excitation energies of \subt{A}-dominated transitions that are consistent through second order. 
The explicit expressions for the terms are given by
\begin{align}
\text{I}&=\langle 0_\subm{A} | [{\mathbf{Q}}_\subm{A},[\hat{\mathcal{H}}_\subm{A}+\hat{\mathcal{V}}^{\text{int}},{\mathbf{Q}}_\subm{A}^{\dagger}]]|0_\subm{A}\rangle\ \text{,}\label{eq:I}\\[0.05in]
\text{II}&= \mathbf{V}_{\hat{\mathcal{V}}^\subm{A}_\mathbf{r}}C_{\mathbf{r},\mathbf{r}'}^{\subm{B},(0)}(\omega)\mathbf{V}_{\hat{\mathcal{V}}^\subm{A}_{\mathbf{r}'}}^{\dagger}\ \text{,}\label{eq:II}
\end{align}
using Eq.~\eqref{eq:int_operator_perturbation} for a single environment subsystem and implying that $|0_\subm{A}\rangle$ has been derived from the effective Hamiltonian including  Eq.~\eqref{eq:int_operator_perturbation} rather than the full operator in Eq.~\eqref{eq:eff_Hamiltonian}. The metric matrices retain their structures in Eq.~\eqref{eq:matrices}.
Based on the chosen truncation, the expression for the effective property gradient for subsystem \subt{A} becomes
\begin{align}\label{eq:propertygradient_pert}
\widetilde{\mathbf{V}}_{\subm{A},\alpha}^{\omega}=&\mathbf{V}_{\subm{A},\alpha}^{\omega(2)}-\mathbf{E}_\subm{AB}^{[2](1)}(\mathbf{E}_\subm{B}^{[2](0)}-\omega\mathbf{S}_\subm{B}^{[2](0)})^{-1}\mathbf{V}_{\subm{B},\alpha}^{\omega(0)}\ \text{.}
\end{align}
The analogy to SOPPA is thus imperfect, since keeping only the zeroth-order correction to the \subt{B} part of the property gradients means that the effective transition moments and thereby the linear response function are not consistent through second order.

\begin{figure*}
  \includegraphics[width=1.0\textwidth]{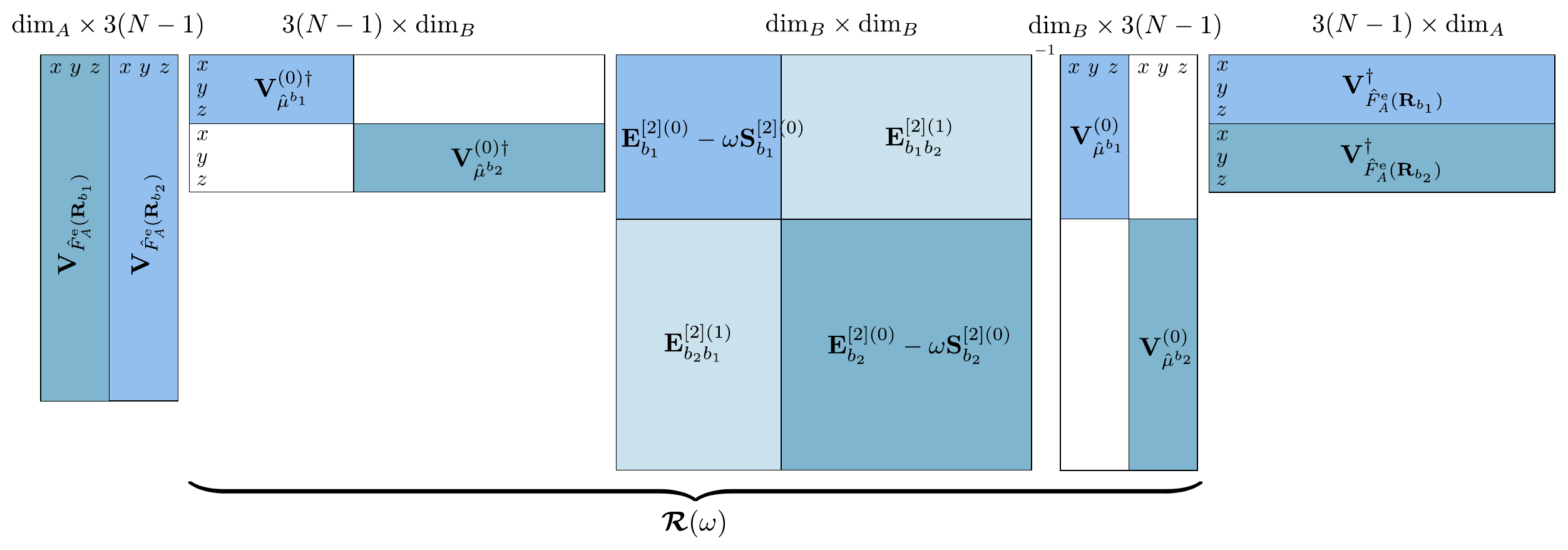}
  \caption{Structure of the PE analog of Eq.~\eqref{eq:II} for an environment with two ($N-1=2$) environment subsystems $b_1$ and $b_2$. White blocks are zero, while colored blocks represent nonzero intra- and intersubsystem blocks. Upon repeated use of Eq.~\eqref{eq:woodbury2}, the expression can be written in PE terminology in terms of the relay matrix $\bm{\mathcal{R}}(\omega)$.}
\label{fig:matrixstructure}
\end{figure*}

To arrive at the PE model, we further need to decompose the environment into individual subsystems and invoke a truncated multipole representation of the interaction operator with respect to the environment subsystems. For practical feasibility but without theoretical justification, the lowest-order approximation invoked for the combined environment is also employed for all individual subsystems constituting the environment, meaning that the ground-state polarization among the environment subsystems, otherwise implied in $\mathbf{E}^{[2](0)}_\subm{B}$, is neglected. Taking the simplest two-subsystem environment as an example, the structure of the PE analog of II in Eq.~\eqref{eq:II} then takes the form illustrated in Fig.~\ref{fig:matrixstructure}. 
In particular, upon rewrite of the matrix resolvent according to Eq.~\eqref{eq:matrixlemma}, contraction with the property gradients and repeated use of Eq.~\eqref{eq:woodbury2} on the resulting subsystem blocks, we recognize the series expansions of the corresponding blocks in the relay matrix. For instance, the first block can be rewritten as
\begin{align}
\mathbf{V}_{\hat{\mu}^{b_1}}^{(0)\dagger}&\Bigl((\mathbf{E}_{b_1}^{[2](0)}-\omega\mathbf{S}_{b_1}^{[2](0)})^{-1}-\mathbf{E}_{b_1b_2}^{[2](1)}(\mathbf{E}_{b_2}^{[2](0)}-\omega\mathbf{S}_{b_2}^{[2](0)})^{-1}\mathbf{E}_{b_2b_1}^{[2](0)}\Bigr)^{-1}\mathbf{V}_{\hat{\mu}	^{b_1}}^{(0)}\notag\\
&=\Bigl(\bm{\alpha}_{b_1}(\omega)^{-1}+\mathbf{T}^{(2)}_{b_1b_2}\bm{\alpha}_{b_2}(\omega)\mathbf{T}^{(2)}_{b_2b_1}\Bigr)^{-1}\ \text{,}
\end{align}
that is, in terms of the relay matrix in Eq.~\eqref{eq:relaymatrix}. Rewriting the second term of Eq.~\eqref{eq:propertygradient_pert} in a similar manner, we finally obtain the PE analogs of the effective electronic Hessian and effective property gradient defined in Ref.~\citenum{list2016local}:
\begin{align}
\widetilde{\mathbf{E}}_\subm{A}^{[2]}&\stackrel{{\text{PE}}}{=}\langle 0_\subm{A}| [{\mathbf{Q}}_\subm{A},[\hat{\mathcal{H}}_\subm{A}+\hat{v}_\text{PE},{\mathbf{Q}}_\subm{A}^{\dagger}]]| 0_\subm{A}\rangle\notag\\
&-\sum_{s,s'=1}^S\langle 0_\subm{A}| [{\mathbf{Q}}_\subm{A},\hat{\mathbf{F}}^{\text{e}}_{\subm{A}}(\mathbf{R}_s)]| 0_\subm{A}\rangle \bm{\mathcal{R}}_{ss'}(\omega)\langle 0_\subm{A}| [{\mathbf{Q}}^{\dagger},\hat{\mathbf{F}}^{\text{e}}_{\subm{A}}(\mathbf{R}_{s'})]| 0_\subm{A}\rangle\ \text{,} \label{eq:hessian_PE}\\
\widetilde{\mathbf{V}}_{\subm{A},\alpha}^{\omega}&\stackrel{{\text{PE}}}{=}\langle 0_{\subm{A}}| [{\mathbf{Q}}_\subm{A},\hat{V}^\omega_{\subm{A},\alpha}]| 0_\subm{A}\rangle -\sum_{s,s'=1}^S\langle 0_{\subm{A}}| [{\mathbf{Q}}_\subm{A},\hat{\mathbf{F}}^{\text{e}}_{\subm{A}}(\mathbf{R}_s)]| 0_\subm{A}\rangle\bm{\mathcal{R}}_{ss'}(\omega)\mathbf{e}_{\alpha}\ \text{,}\label{eq:propertygradient_PE}
\end{align}
taking $\hat{\mu}_{\alpha}$ as perturbation and using $\mathbf{e}_\alpha$ to denote a unit vector in the Cartesian $\alpha$ direction. The above expressions define the environmental effects included in the PE model within a linear response framework. Specifically, Eq.~\eqref{eq:hessian_PE} defines the static (through $\hat{v}_{\text{PE}}$) and dynamic reaction field effects (second term), while the second term of  Eq.~\eqref{eq:propertygradient_PE} defines the effective external field effect.\cite{list2016local}
In this way, we have shown the transition from a full quantum-mechanical description of the linear response of the combined system to the differentiated subsystem treatment in the PE model. Based on the insight from the theoretical analysis in Sec.~\ref{sec:compositesystems}, these effective quantities can then be
used in the first term of the SD (NSD) to obtain the subsystem \subt{A} contributions to molecular (transition) properties.

A commonly adopted possibility to further reduce the computational complexity of the calculations is to assume frequency-independent environment subsystems, corresponding to imposing 
$\mathbf{S}_{\subm{B}}^{[2]}=\mathbf{0}$. 
This zero-frequency (ZF) approximation offers significant simplifications: (\textit{i}) the nonlinearity of the effective electronic Hessian is lost such that Eq.~\eqref{eq:eff_eigenvalueeq} reduces to a standard generalized eigenvalue problem, (\textit{ii}) the renormalization factor, otherwise needed in Eq.~\eqref{eq:symmetric_M}, becomes unity because the excitation is restricted to subsystem \subt{A} in this approximation, and (\textit{iii}) the additional zeroth-order poles in the first term of Eq.~\eqref{LRF-complex-symmetric} corresponding to the FP approximation are removed.
According to previous studies,\cite{harczuk2015frequency,norby2016assessing} the zero-frequency limit is a good approximation at off-resonant and optical frequencies because of the larger excitation energy 
where the frequency dispersion in the environment subsystems is typically small. In such cases, the dynamical response of the environment on the excitations in subsystem \subt{A} is captured reasonably well by the static limit.

\section{Numerical illustration}\label{sec:results}
\begin{figure}
  \includegraphics[width=1.0\columnwidth]{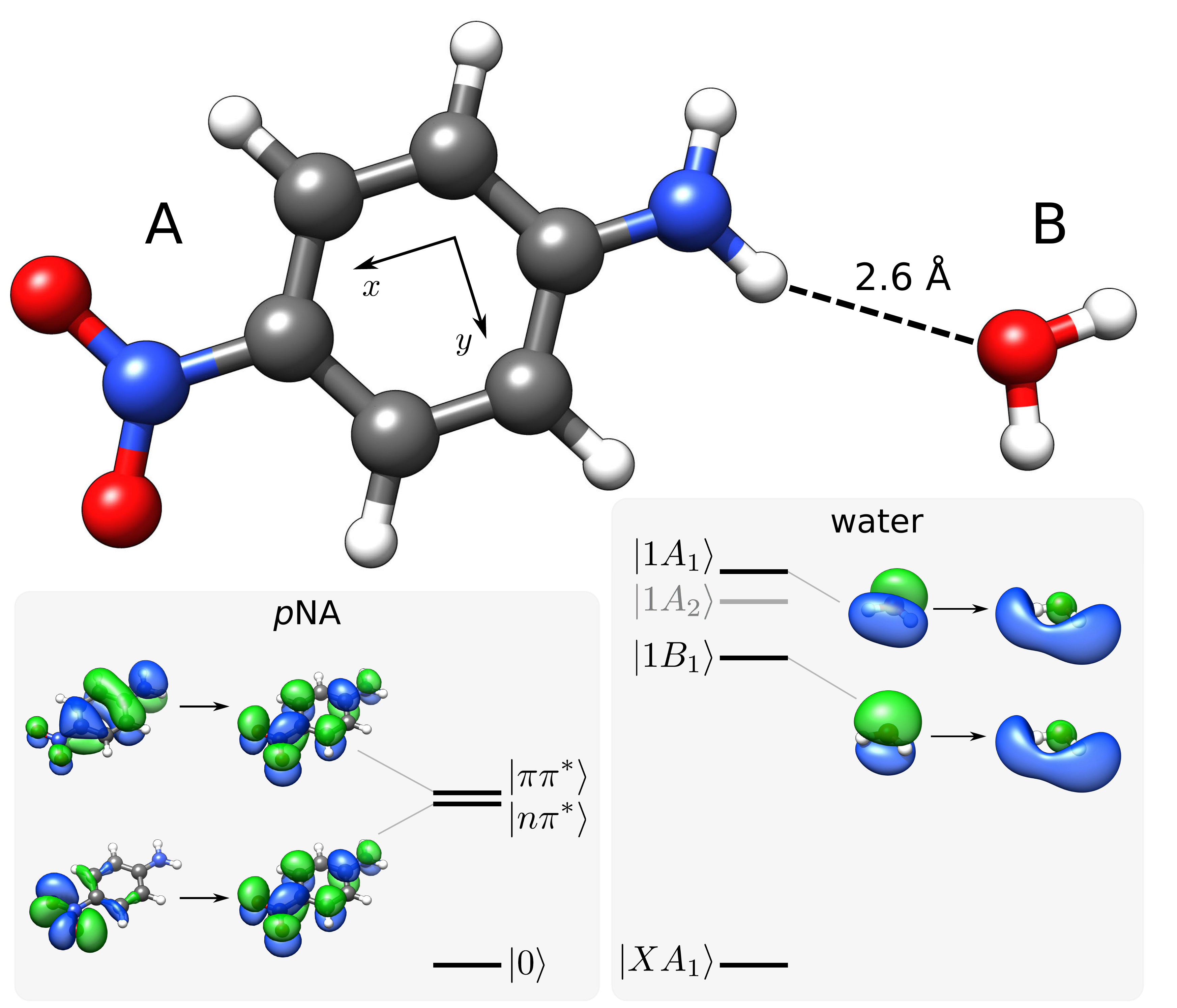}
  \caption{The six-level-model used for the \textit{p}NA--water complex with isosurfaces of the orbitals mainly involved in the considered transitions (shown in black).}
\label{fig:SLM}
\end{figure}

To illustrate the basic features of the response of a combined system to a perturbing external field and the importance of the various intersubsystem interactions, we will in this section perform a numerical inspection of the working expressions presented in Sec.~\ref{sec:theory}. 
For this purpose, we consider a simplified six-level-model (SLM) for a \textit{para}-nitroaniline(\textit{p}NA)--water complex and its linear response to a uniform electric-field perturbation. In addition to the respective ground states, the SLM includes also the first and second singlet excited states for \textit{p}NA---these are the $n\pi^*$ state and the intramolecular amino-to-nitro charge-transfer transition referred to as $\pi\pi^*$---and the first and third singlet excited states for water---these are states $1B_1$ and $1A_1$, respectively, using the symmetry labels referring to the irreducible representations of the $C_{2v}$ point group of the parent molecule. The manifold of states in the SLM is depicted in Fig.~\ref{fig:SLM}, defining \textit{p}NA and water as subsystem \subt{A} and \subt{B}, respectively. The set of monomer parameters and electronic couplings reported in Tables \ref{tab:data} and \ref{tab:electronic_couplings} have been obtained at the TD-DFT level of theory employing CAM-B3LYP\cite{yanai2004new}/aug-cc-pVDZ\cite{dunning} in the presence of the ground-state-frozen embedding potential of the other system (i.e., corresponding to the FP approximation), as obtained from a PE calculation. In other words, the second term in Eq.~\eqref{eq:hessian_PE} is excluded in the response calculation. The embedding potentials consisted of atom-centered permanent electric multipoles up to quadrupoles and anisotropic electric dipole--dipole polarizabilities and were computed according to the LoProp\cite{loprop} scheme using DALTON and the Loprop-for-Dalton Python script.\cite{Vahtras:13276}
The calculations were performed using a development version of the DALTON program\cite{dalton,dalton2} that contains the implementation of electronic couplings between transition densities.\cite{steinmann2015electronic} The nuclear configuration of the complex has been taken from a molecular dynamics simulation.\cite{sneskov2011scrutinizing} We use this polarized basis as approximate representation for the eigenvectors of the subsystem electronic Hessians, such that the diagonal blocks of the full Hessian are diagonal. As expected from the strengths of the leading-order transition dipoles and their relative orientations (the charge-transfer transition is directed along the $x$-axis), only the electronic coupling between the $\pi\pi^*$ and $1A_1$ states is significant whereas the $n\pi^*$ and $1B_1$ states will essentially be unaffected. However, the state mixing still remains small because their energy difference is significantly larger than their electronic coupling.
Therefore, to better illustrate the different aspects of the response of a combined system, the coupling blocks of the electronic Hessian have been scaled by a factor of 12. 
 The largest absolute intersubsystem coupling element is then equal to 0.0212 a.u., which is smaller than the relevant difference between the excitation energies in the subsystems by a factor of $\sim$9. 

 \begin{table}[h!]
 \caption{The polarized subsystem parameters used in the SLM for the \textit{p}NA--water complex in Fig.~\ref{fig:SLM}. All results are reported in a.u.~and have been obtained at the CAM-B3LYP/aug-cc-pVDZ level of theory in the presence of the ground-state-frozen PE embedding potential 
     of the other subsystem.}
\begin{tabular*}{\columnwidth}{@{\extracolsep{\fill}}ccccc}
  \hline \hline\noalign{\smallskip}
State & $\Delta E$ & $|\mu_x|$ & $|\mu_y|$ & $|\mu_z|$ \\
\noalign{\smallskip}\hline\noalign{\smallskip}  
$|n\pi ^*\rangle$  & 0.14014 & 0.00102	& 0.00017 & 0.00023\\
$|\pi\pi^*\rangle$ & 0.15428 & 2.02274  & 0.00030 & 0.00539\\
$|1B_1\rangle$     & 0.26814 & 0.07302 	& 0.02634 & 0.54089\\
$|1A_1\rangle$     & 0.35027 & 0.42522  & 0.49612 & 0.09183 \\
\noalign{\smallskip}\hline \hline
  \end{tabular*}
        \label{tab:data}
\end{table}

 \begin{table}[h!]
 \caption{Absolute (unscaled) electronic couplings used in the SLM for the \textit{p}NA--water complex in Fig.~\ref{fig:SLM}. Digits in parenthesis refer to negative exponents, {i.e.}, $a(b)=a \times 10^{-b}$. All results are reported in a.u.~and have been obtained at the CAM-B3LYP/aug-cc-pVDZ level of theory in the presence of the ground-state-frozen PE embedding potential of the other subsystem.}
\begin{tabular*}{\columnwidth}{@{\extracolsep{\fill}}ccc}
  \hline \hline\noalign{\smallskip}
pNA$\backslash$water & 
  $|1B_1\rangle$& $|1A_1\rangle$   \\
\noalign{\smallskip}\hline\noalign{\smallskip}  
$|n\pi^*\rangle$ & 2.53(6) & 1.03(6) 	\\
$|\pi\pi^*\rangle$ & 3.82(6)  & 1.76(3) \\
\noalign{\smallskip}\hline \hline
  \end{tabular*}
        \label{tab:electronic_couplings}
\end{table}

\begin{figure}
  \includegraphics[width=\columnwidth]{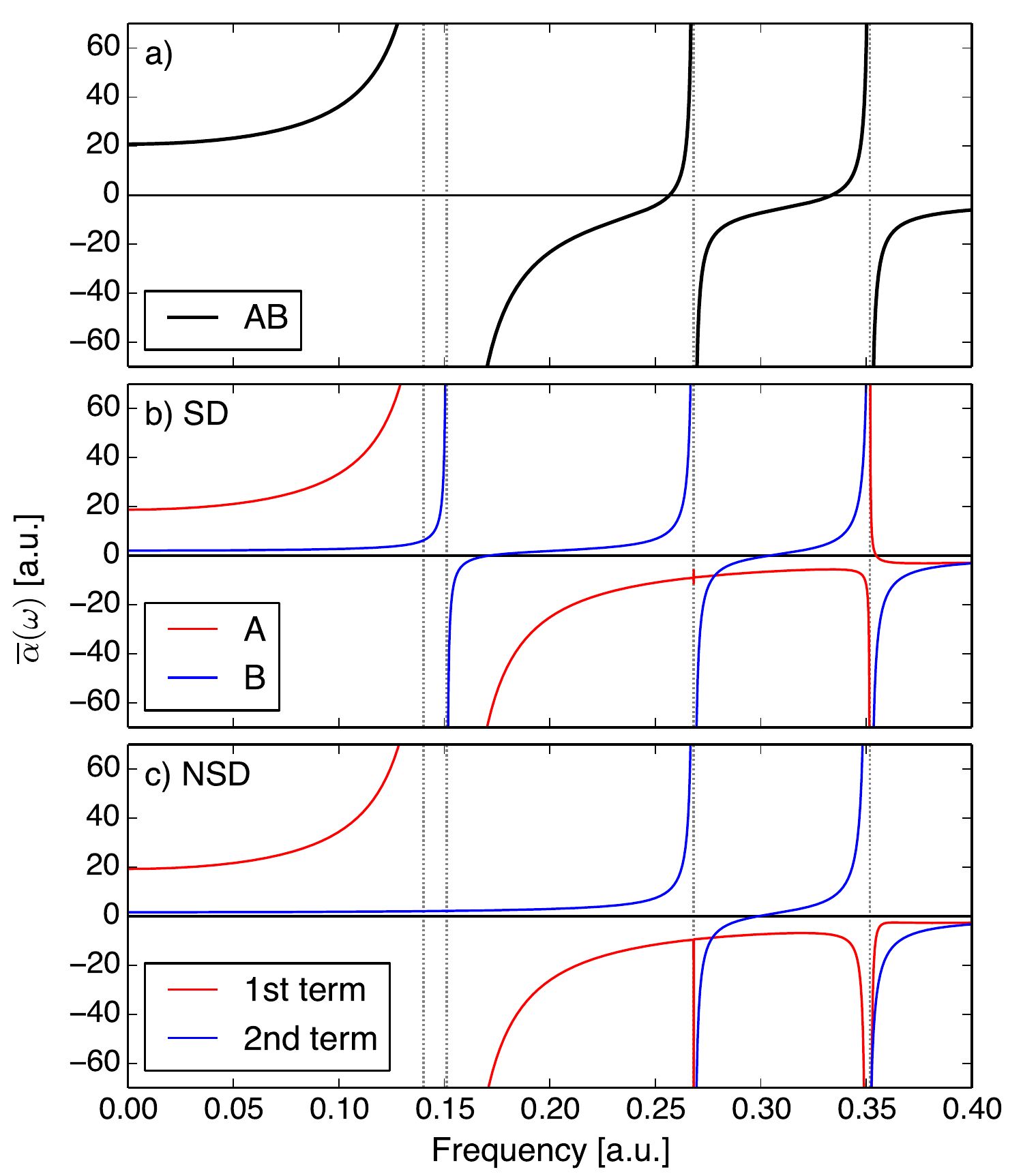}
  \caption{Isotropic electric dipole--dipole polarizability adopting the SLM in Fig.~\ref{fig:SLM}. (a) For the combined system as given by Eq.~\eqref{eq:RTCS_LRfunction_compact}, (b) the symmetric decomposition according to Eq.~\eqref{eq:molprops}, and (c) the nonsymmetric decomposition according to Eq.~\eqref{eq:decomposed}. Vertical dotted lines indicate the resonance frequencies of the combined system.}
\label{fig:dynamic_pol}
\end{figure}

Fig.~\ref{fig:dynamic_pol}a shows the isotropic electric dipole--dipole           polarizability, Eq.~\eqref{eq:RTCS_LRfunction_compact}, of the \textit{p}NA--water complex within the SLM. The excitation energies of the full system (vertical dotted lines) can be identified as the poles of the linear response function. In the frequency region around the two lowest poles ($\omega \approx 0.10-0.20$ a.u.), the isotropic polarizability is dominated by the $\alpha_{xx}$ component of the tensor with a dispersion that is in turn dictated by the charge-transfer transition (second pole). This leads to a seemingly absence of a pole at the first excitation in subsystem \subt{A}, but it is thus merely a consequence of the $n\pi^*$-transition being nearly electric-dipole forbidden and close in energy to the intense $\pi\pi^*$-transition. 
 
\begin{figure}
  \includegraphics[width=\columnwidth]{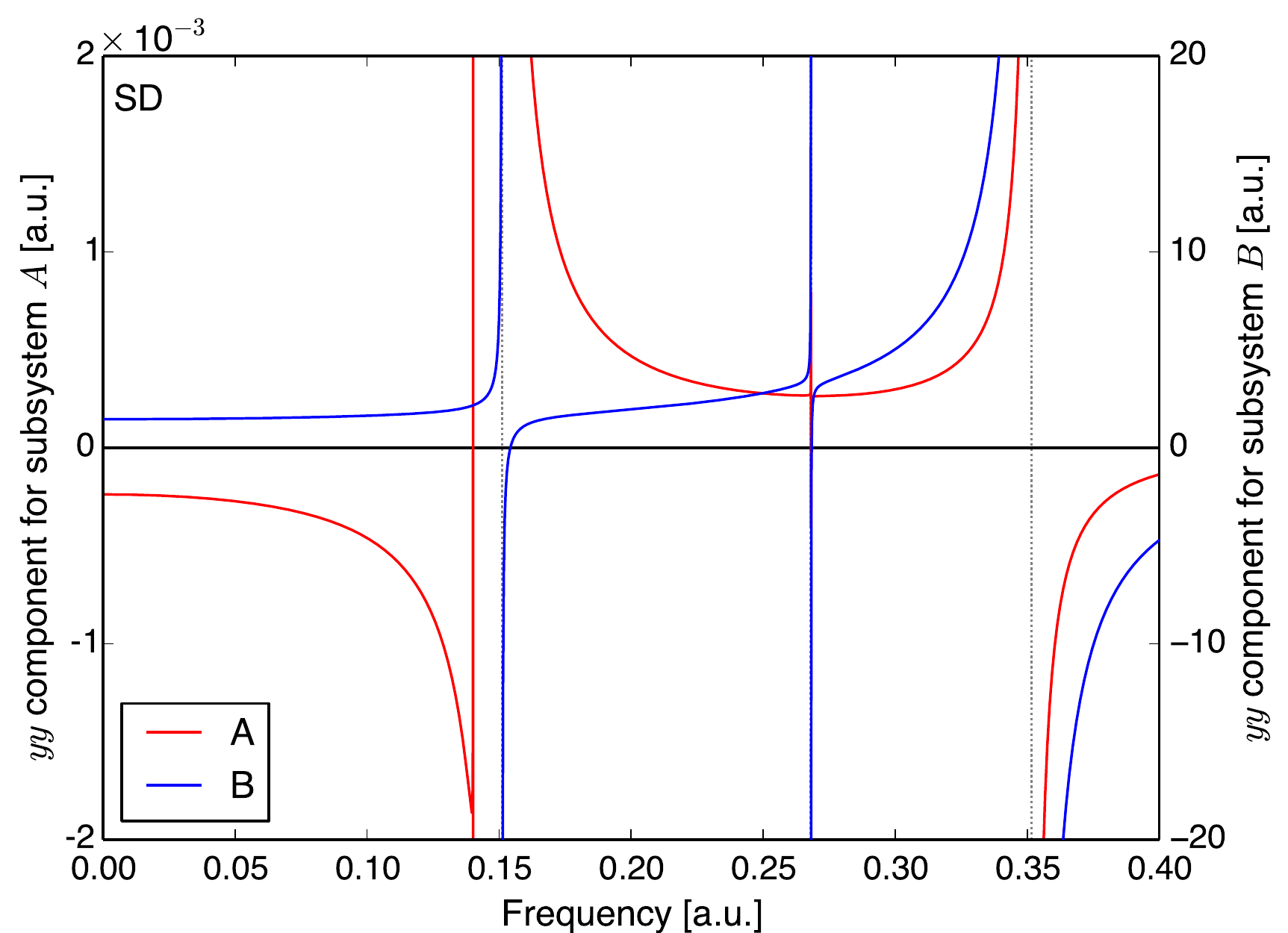}
  \caption{Symmetric subsystem contributions to the $yy$ component of the electric dipole--dipole polarizability of the combined system adopting the SLM in Fig.~\ref{fig:SLM}. Note the different scales used for the two $y$-axes. The vertical dotted lines indicate the resonance frequencies of the combined system.}
\label{fig:yyplot}
\end{figure}

The symmetric decomposition of the polarizability according to Eq.~\eqref{eq:molprops} into subsystem \subt{A} and \subt{B} contributions is shown in Fig.~\ref{fig:dynamic_pol}b. Note that while the diagonal elements of the polarizability of the combined system in the static limit are guaranteed to be positive, the same does not hold for the individual subsystem contributions. As exemplified in Fig.~\ref{fig:yyplot} by the subsystem \subt{A} contribution to $\alpha_{yy}$ (the first term in Eq.~\eqref{eq:molprops}), this situation does indeed occur in the present case. As seen in Eq.~\eqref{eq:effectivegradients}, there are two contributions to the modified property gradient $\widetilde{\mathbf{V}}^\omega_{\subm{A},\beta}$. The contribution from the first term, {i.e.},~$\mathbf{V}^\omega_{\subm{A},\beta}$, is guaranteed to give a positive contribution to $\alpha_{yy}$ for subsystem \subt{A} but, due the large response in subsystem $\subt{B}$, the second term in the effective property gradient becomes dominant and leads to an overall negative value for $\alpha_{yy}(0)$ of subsystem \subt{A}.

Furthermore, we note that both the subsystem \subt{A} and \subt{B} contributions in the symmetric decomposition contain poles at \textit{all} the excitation energies of the combined system, also referred to as the physical excitations. As we discussed previously, this implies that transition moments, in contrast to excitation energies, cannot be determined from any single one of the terms. With the nonsymmetric decomposition (Eq.~\eqref{eq:decomposed}, Fig.~\ref{fig:dynamic_pol}c), on the other hand, all the physical poles are collected in the first term, and it is thus the proper choice when determining residues for the transitions mainly located in subsystem \subt{A}. In addition to the physical poles, however, there are unphysical zeroth-order poles in both the first and second terms of Eq.~\eqref{eq:decomposed}. Specifically, they contain poles at the excitation energies of subsystem \subt{B} within the FP approximation, i.e., where the environment polarization is fixed during the response calculation. This inclusion of both physical and unphysical poles in the first term of Eq.~\eqref{eq:decomposed} becomes particularly apparent in the frequency region close to the fourth excitation in Fig.~\ref{fig:dynamic_pol}c (see red solid line at $\omega \approx 0.35$ a.u.).

The excitation energies and associated one-photon transition strengths are reported in Table \ref{tab:results} for the two lowest transitions in the model system, {i.e.}, those predominantly localized in subsystem \subt{A}.
 \begin{table*}[ht!]
	\caption{Excitation energies and one-photon transition strengths for the two lowest-lying (i.e.,~subsystem \subt{A}-dominated) electronic transitions in the composite system as computed from the full treatment as well as the nonsymmetric decomposition (NSD), including and excluding various contributions. All results are reported in a.u.~and digits in parenthesis refer to negative exponents, {i.e.}, $a(b)=a \times 10^{-b}$. vac: vacuum, FP: frozen ground-state environment polarization, renorm: renormalization, EEF: effective external field, ZF: zero-frequency approximation for the environment.}
	\begin{tabular*}{\textwidth}{@{\extracolsep{\fill}}ccccccccc}
		\hline \hline\noalign{\smallskip}
		& \multicolumn{4}{c}{$|1_{\subm{AB}}\rangle$} &   \multicolumn{4}{c}{$|2_{\subm{AB}}\rangle$}  \\
		\noalign{\smallskip}
		\cline{2-5}\cline{6-9}
		\noalign{\smallskip}
		Model & $\Delta E$ & $|\mu_x|^2$ & $|\mu_y|^2$ & $|\mu_z|^2$ & $\Delta E$ & $|\mu_x|^2$& $|\mu_y|^2$ & $|\mu_z|^2$\\
		\noalign{\smallskip}\hline\noalign{\smallskip}  
		vac               & 0.13950 & 6.0205(7) & 3.3384(8) & 6.6642(9) & 0.15625 & 4.0014 & 6.5832(7) & 4.3947(5) \\
		FP             & 0.14014 & 1.4022(6) & 3.0284(8) & 5.1435(8) & 0.15428 & 4.0915 & 8.8700(8) & 2.9025(5) \\
		\noalign{\smallskip}\hline\noalign{\smallskip} 
		Full system             & \multirow{5}{*}{0.14014} & 1.9202(6) & 4.8179(8) & 2.2571(9) & \multirow{5}{*}{0.15110} & 4.3996 & 5.4416(3) & 3.5539(4)\\  
		NSD (Eq.~\eqref{eq:symmetric_M})                &         & 1.9202(6) & 4.8179(8) & 2.2571(9) &  & 4.3996 & 5.4416(3) & 3.5539(4)\\
		NSD($-$renorm)     &         & 1.9202(6) & 4.8179(8) & 2.2571(9) &         & 4.4423 & 5.4944(3) & 3.5884(4) \\
		NSD($-$EEF)               &         & 1.8561(6) & 3.0263(8) & 5.1047(8) &         & 4.1373 & 8.9645(8) & 2.9350(5) \\
		NSD($-$(renorm/EEF))      &         & 1.8561(6) & 3.0263(8) & 5.1047(8) &         & 4.1774 & 9.0515(8) & 2.9635(5) \\
		NSD(+ZF)         & 0.14014 & 1.7309(6) & 4.4586(8) & 9.1521(9) & 0.15170 & 4.3745 & 3.6180(3) & 2.7005(4) \\
		\noalign{\smallskip}\hline \hline
	\end{tabular*}
	\label{tab:results}
\end{table*}
 First, the results provide a clear evidence for the equivalence between the properties obtained from the decomposed subsystem expressions in
Eqs.~\eqref{eq:eff_eigenvalueeq} and \eqref{eq:symmetric_M} and from the consideration of the full system expressed in terms of Eq.~\eqref{eq:RTCS_eigenvalue_equation}. 
We further consider various approximate models that are defined according to what terms are retained in the response expressions. The impact of the renormalization factor defined in
Eq.~\eqref{eq:normalization} and appearing in Eq.~\eqref{eq:symmetric_M} depends on the degree of delocalization of the given transition, and its neglect (denoted by "$-$renorm" in Table~\ref{tab:results}) would lead to an overestimation of transition strengths. The effective field strengths experienced by subsystem \subt{A} in the presence of subsystem \subt{B} can be different for different directions as a consequence of the anisotropy of the polarizability of \subt{B} and the relative orientation of the interacting subsystems, as seen by comparing to the assumption that only subsystem \subt{A} interacts with the external field (denoted by "$-$EEF" in Table~\ref{tab:results}). For excitations in subsystem \subt{A} far from resonances in \subt{B} (the environment), as in the present case, the impact of renormalization is smaller than that of the EEF effect, as expected from the relative distance dependency of the two contributions.

\begin{figure}
 \includegraphics[width=\columnwidth]{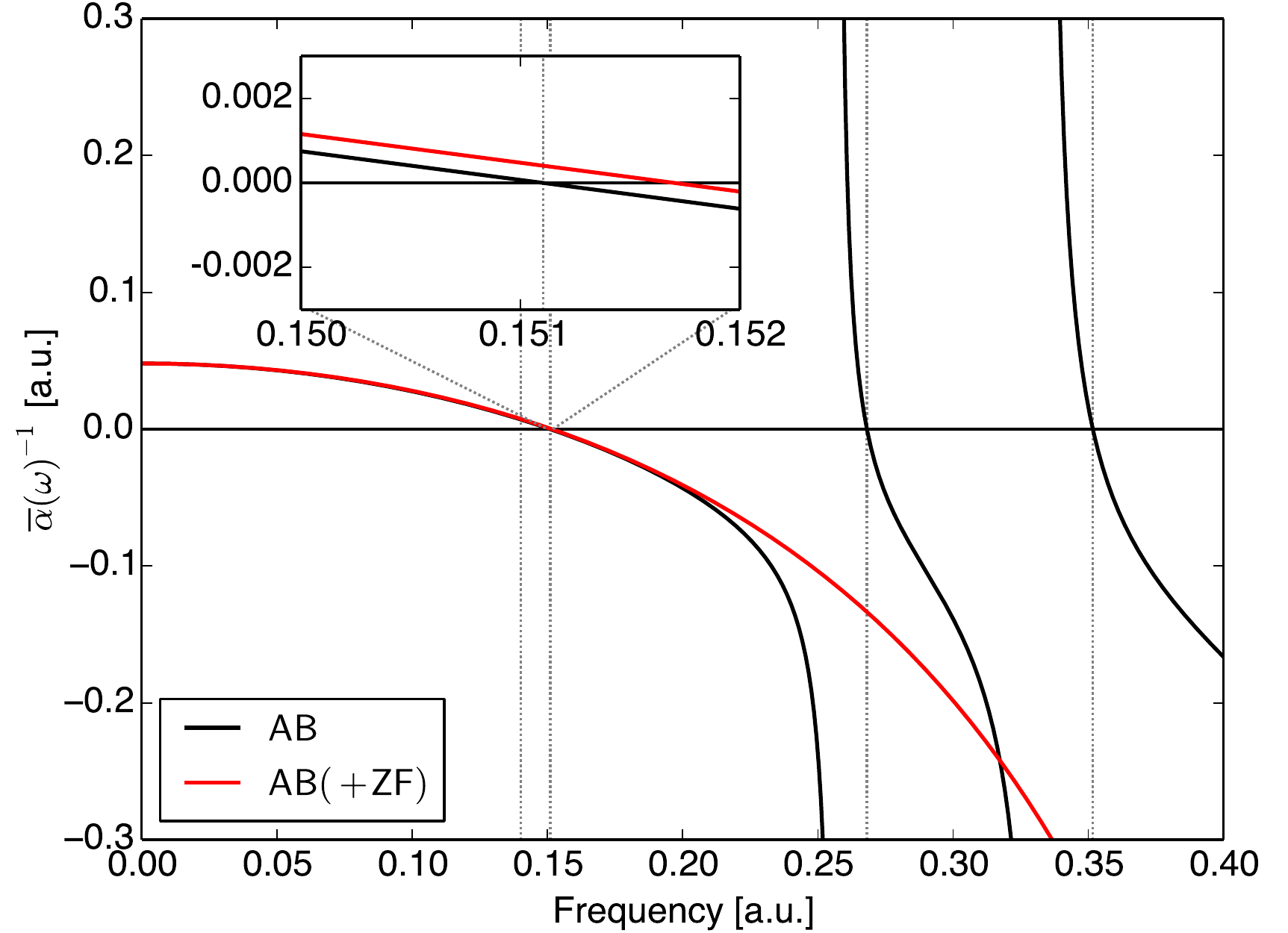}
  \caption{Frequency-dependency of the inverse of the isotropic electric dipole--dipole polarizability of the \textit{p}NA--water complex within the SLM when including the dynamic response of water compared to its zero-frequency limit. This approximation reduces the number of poles to the number of excitations in subsystem \subt{A} and blue-shifts the excitation energies (see inset).}
\label{fig:zerofreq}
\end{figure}

As discussed in Sec.~\ref{sec:PE_response}, the various polarizable embedding models typically assume the effective electronic Hessian of subsystem \subt{A} to be frequency-independent, thereby losing the nonlinearity of the linear response equations and the poles associated with excitations predominantly in subsystem \subt{B}. The zero-frequency limit of the inverse of the isotropic polarizability of the combined system is illustrated in Fig.~\ref{fig:zerofreq} along with the full frequency-dependent counterpart. Associated excitation energies and transition moments are given in Table \ref{tab:results}. As anticipated from the weak dispersion of the real polarizability for subsystem \subt{B} at the resonance frequencies in subsystem \subt{A} (see Fig.~\ref{fig:dynamic_pol}c), the neglect of transitions in subsystem \subt{B} only leads to a small changes in the transition properties of the \subt{A}-dominated excitations. 
It is noted that the ZF approximation is analogous to the familiar adiabatic approximation in TD-DFT where the exchange--correlation kernel is assumed to be frequency-independent and hence leads to a linear eigenvalue problem with solutions only at Kohn--Sham one-electron excitations.\cite{maitra2004double, cave2004dressed, casida2015many} Double- and higher-electron excitations and their effects on the Kohn--Sham single excitations are thus neglected in adiabatic TD-DFT,\cite{caballero2013comparison} as are the excitations in subsystem \subt{B} in our case.

\begin{figure}[h!]
  \includegraphics[width=\columnwidth]{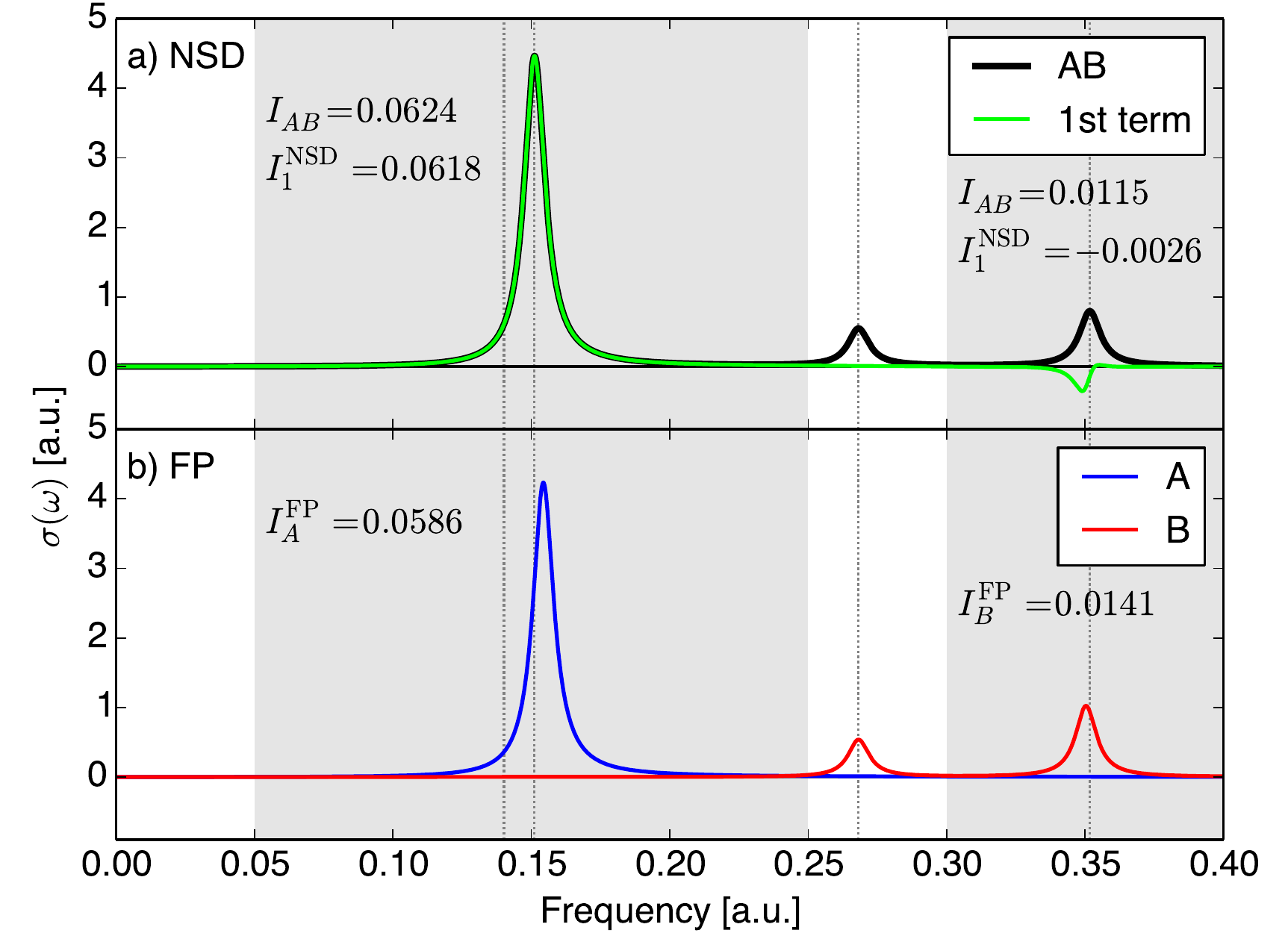}
  \caption{The linear absorption cross section for \textit{p}NA--water complex within the SLM. (a) 
  Integrated absorption cross sections (in a.u.) are reported for the intervals indicated by the gray-shaded areas. A common damping parameter of $4.5566\times 10^{-3}$ a.u.~was used for all transitions. The vertical dotted lines indicate the resonance frequencies of the combined system.}
\label{fig:imag_xx_intensityborrowing}
\end{figure}

As discussed in Sec.~\ref{sec:damped}, the coupling of subsystem excitations gives rise to intensity borrowing. To illustrate this effect, we report in Fig.~\ref{fig:imag_xx_intensityborrowing}a the linear absorption cross section defined in Eq.~\eqref{eq:cross-section} for the full system together with that associated with subsystem \subt{A}, {i.e.}, the first term of Eq.~\eqref{LRF-complex-symmetric}. In Fig.~\ref{fig:imag_xx_intensityborrowing}b, the corresponding cross sections for the subsystems within the FP approximation are reported. First, we note that upon integrating the absorption cross section for the full system (black solid line in Fig.~\ref{fig:imag_xx_intensityborrowing}a) across the full frequency range, we obtain a reference value of $I_{\subm{AB}}=0.0829$ a.u. Due to the very limited description of the electronic structure of the \textit{p}NA--water complex by means of the SLM, this value is far from the exact value of 11.81 a.u.~as obtained from the conservation law in Eq.~\eqref{eq:conservationlaw} for a system with 82 electrons. But this discrepancy is of no concern here since the key point is that the value for the integrated cross section is identical to the corresponding summed result for the two  subsystems within the FP approximation ($I_\subm{A}^{\text{FP}} = 0.0606$ and $I_\subm{B}^{\text{FP}} = 0.0223$ a.u., respectively).

Let us now pass to examine the nonsymmetric decomposition of the interacting system with individual cross sections $\sigma_1^\mathrm{NSD}(\omega)$ and $\sigma_2^\mathrm{NSD}(\omega)$ for the two terms in Eq.~\eqref{LRF-complex-symmetric}, respectively. These two individual cross sections obey their own respective conservation laws as expressed in Eq.~\eqref{eq:conservationlaw2}. Since the nonsymmetric decomposition is made such that the second term corresponds exactly to subsystem \subt{B} within the FP approximation, the cross section $\sigma_2^\mathrm{NSD}(\omega)$ is identical to the red solid line in Fig.~\ref{fig:imag_xx_intensityborrowing}b.

Finally, we include in Fig.~\ref{fig:imag_xx_intensityborrowing} also the partially integrated absorption cross section for two separate finite energy intervals, shown as gray-shaded areas in the figure. In agreement with the increase in transition strength found in Table \ref{tab:results}, the lowest band dominated by the $\pi\pi ^*$ state is intensified upon coupling the excitations in the two subsystems. This gain in intensity in one part of the spectrum is counteracted by a reduction of the intensity in another part and,
in the nonsymmetric decomposition, this will be seen only in $\sigma_1^\mathrm{NSD}(\omega)$. Since the coupling of the lowest subsystem \subt{A}-dominated band (second transition) is only effective to the subsystem \subt{B}-dominated $1A_1$-band (fourth transition), there will be an intensity borrowing from the latter to the former. As a consequence, $\sigma_1^\mathrm{NSD}(\omega)$ will take on negative values in the region of the fourth transition ($\omega \approx 0.35$ a.u.) as seen in Fig.~\ref{fig:imag_xx_intensityborrowing}a (green solid line). Accordingly, in frequency regions dominated by transitions in subsystem \subt{A}, the calculation of $\sigma_1^\mathrm{NSD}(\omega)$ is to be associated with an absorption spectrum, whereas this cannot be readily done in frequency regions that includes transitions in subsystem \subt{B}.

\section{Summary and Conclusions}
In this work, we have provided a rigorous derivation of the various contributing terms appearing in linear response theory of a quantum molecular system embedded in a polarizable environment. 
The origin of the three distinct types of mechanisms for the environmental effects within polarizable embedding---the static and dynamic environment responses entering the intra- and intersubsystem blocks of the electronic Hessian, respectively, as well as the effective external field effect modifying the property gradient---follow in a straightforward manner from a subsystem partitioning of a quantum-mechanical direct-product treatment of the response of the entire system. In particular, the effective external field effect is a consequence of the direct interaction between the environment and the probing external field and leads to the definition of effective subsystem properties.

A crucial point is what decomposed form of the response function that ought to be used in a given context. In the present theoretical analysis, we have demonstrated the basic features of two alternative subsystem decompositions, which clarify such discussions and highlight potential issues in defining subsystem contributions to response and transition properties.

Our first decomposition provided in Eq.~\eqref{eq:molprops} treats subsystems on an equal footing and results in two linear response function terms that express the permutation of subsystem indices. For that reason, we have denoted this decomposition scheme as {symmetric}.
This is the natural choice for computing subsystem contributions to molecular properties, but, as already shown in the framework of subsystem DFT,\cite{pavanello2013subsystem} the coupling of the subsystems manifests itself in that each subsystem contribution contains the poles of the entire system---a feature that emphasizes the approximate nature of our picture of "localized" transitions. As a result that is made clear in the present work, transition strengths for a subsystem-dominated excitation cannot be found from a residue analysis of the symmetric subsystem linear response function.

Our second decomposition provided in Eq.~\eqref{eq:decomposed} treats, on the other hand, subsystems on an unequal footing and results in two linear response function terms that both are symmetric with respect to left and right property gradients. This is an unnatural choice for the development of polarizable embedding models due to the fact that the first term of the response function (describing the response properties of the chromophore of interest) contains poles not only from all transitions in the fully interacting system but also poles from the ground-state polarized, but otherwise uncoupled, environment. It is demonstrated that the linear response function in the nonsymmetric decomposition lends itself to a determination of transition strengths but only after taking proper account of renormalization of the subsystem excitation vectors. However, the renormalization factor is not accessible within the framework of polarizable embedding because it cannot be rewritten in terms of the response kernel of the environment. This complication can be avoided in practice by assuming the static limit for the environment response or by turning to the framework of complex linear response theory. We have shown that the integrated absorption cross sections of the two terms in the symmetric decomposition of the linear response function are preserved independently of each another and, as a consequence, it is demonstrated (also in the numerical example of the water--\textit{p}NA dimer) that subsystem absorption cross sections (and likewise oscillator strengths) will take on negative values if intersubsystem intensity borrowing takes place.

\begin{acknowledgments}
N.\ H.\ L.\ thanks J.~Oddershede (University of Southern Denmark, Odense, Denmark) for helpful discussions, and the Carlsberg Foundation for a postdoctoral fellowship (Grant No.~CF15-0792). P.\ N.\ acknowledges financial support from the Swedish Research Council (Grant No.~621-2014-4646). J.\ K.\ thanks the Danish Council for Independent Research (the Sapere Aude program) and the Villum Foundation for financial support. 
Computation/simulation for the work described in this paper has been 
supported by the DeIC National HPC Center, University of Southern Denmark (SDU). 
\end{acknowledgments}

 \renewcommand{\theequation}{A.\arabic{equation}}    
  \appendix 
  \section{Derivation of Eq.~\eqref{eq:conservationlaw}}\label{app:conservationlaw}
  In this appendix, we show how to derive Eq.~\eqref{eq:conservationlaw}. 
  The diagonal representation of the complex linear response function is
  \begin{align}
 - \langle\langle \hat{\mu}_{\alpha}^{-\omega};\hat{\mu}_{\beta}^{\omega}\rangle\rangle=\sum_{n>0}\Biggl[\frac{T_{\alpha\beta}^{0n}}{\omega_n-(\omega+i\gamma)}+\frac{T_{\beta\alpha}^{0n}}{\omega_n+(\omega+i\gamma)}\Biggl].
  \end{align}
  where we have used the definition of the transition strengths in Eq.~\eqref{eq:res} as well as the relation between eigenvectors for paired eigenvalues, i.e., $\mathbf{X}_n=\begin{bmatrix}
  \mathbf{X}_n^1 & \mathbf{X}_n^2\end{bmatrix}$ and $\mathbf{X}_{-n}=\begin{bmatrix}
    \mathbf{X}_n^{2*} & \mathbf{X}_n^{1*}\end{bmatrix}$.\cite{olsen1985linear} Introducing the dispersion and absorption lineshape functions
  \begin{align}
\mathcal{D}_n(\pm\omega)=& \frac{\omega_n\mp\omega}{(\omega_n\mp\omega)^2+\gamma^2}\ \text{,}\\
\mathcal{A}_n(\pm\omega)=& \frac{\gamma}{(\omega_n\mp\omega)^2+\gamma^2}\ \text{,}
  \end{align}
  the real and imaginary components of the complex linear response function can be written as
\begin{align}
-\text{Re}[\langle\langle \hat{\mu}_{\alpha}^{-\omega};\hat{\mu}_{\beta}^{\omega}\rangle\rangle]=&
\sum_{n>0}\Bigl[{T_{\alpha\beta}^{0n}}\mathcal{D}_n(\omega)+T_{\beta\alpha}^{0n}\mathcal{D}_n(-\omega)\Bigl],\\
-\text{Im}[\langle\langle \hat{\mu}_{\alpha}^{-\omega};\hat{\mu}_{\beta}^{\omega}\rangle\rangle]=&
\sum_{n>0}\Bigl[{T_{\alpha\beta}^{0n}}\mathcal{A}_n(\omega)-{T_{\beta\alpha}^{0n}}\mathcal{A}_n(-\omega)\Bigl],
\end{align}

The integrated absorption cross section can therefore be written as
\begin{align}
I=&\int_0^\infty \sigma(\omega)\ \text{d}\omega = \frac{1}{2}\int_{-\infty}^\infty \sigma(\omega)\ \text{d}\omega\ \notag\\[0.1in]
=&\frac{1}{2\epsilon_0c_0} \sum_{n>0}\frac{T_{\alpha\alpha}^{0n}}{3}\int_{-\infty}^\infty \omega 
\Biggl[\frac{\gamma}{(\omega_n-\omega)^2+\gamma^2}-\frac{\gamma}{(\omega_n+\omega)^2+\gamma^2}\Biggl]
\ \text{d}\omega \ \text{,}
\end{align}
where we have used that $\sigma(\omega)$ is an even function. 
Each term in the square bracket can be written as a weighted sum of the dispersion and absorption lineshape functions according to
\begin{align}
\frac{\gamma\omega}{(\omega_n\mp\omega)^2+\gamma^2}=&\mp\gamma\mathcal{D}_n(\pm\omega)\pm\omega_n\mathcal{A}_n(\pm\omega)\ \text{.}
\end{align}
Since $\mathcal{D}_n(\pm\omega)$ is an odd function around $\pm\omega_n$, its integral over the entire frequency range vanishes. Therefore, only $\mathcal{A}_n(\pm\omega)$, which apart from a factor is a Lorentzian function, contributes
\begin{align}
I=&\frac{1}{2\epsilon_0c_0} \sum_{n>0}\frac{\omega_n}{3}T_{\alpha\alpha}^{0n}\int_{-\infty}^\infty 
\bigl[\mathcal{A}_n(\omega)+\mathcal{A}_n(-\omega)\bigl]\ \text{d}\omega\ \notag\\
=&\frac{\pi}{2\epsilon_0c_0}\sum_{n>0}\frac{2\omega_n}{3}T_{\alpha\alpha}^{0n}\notag\\
=&\frac{\pi}{2\epsilon_0c_0}\sum_{n>0}f_{0n}\ \text{.}
\end{align}

\section{Derivation of Eqs.~\eqref{eq:induction_operator_expanded_2} and \eqref{eq:induction_operator_expanded}}\label{app:inductionoperator}  

In this appendix, we shall derive the two multipole-expanded representations of the induction operator given by the first and second equality in Eq.~\eqref{eq:induction_operator_expanded}. The corresponding nonexpanded representations of the operator were given in Eqs.~\eqref{eq:operators2} and \eqref{eq:induction_operator}, respectively, but to ease the discussion, we reiterate both expressions here 
\begin{align}
\hat{\mathcal{V}}^{\text{ind}} =& \sum_{b\in\subt{B}}\int\hat{\rho}^{\text{e}}_{\subm{A}}(\mathbf{r})\langle\hat{\mathcal{V}}_{b}(\mathbf{r})\rangle_{0_{b}}^{(1)}\,\text{d}\mathbf{r}\label{eq:induction_operator1}\\
=& - \sum_{b\in\subm{B}}\int\hat{\rho}_{\subm{A}}^{{e}}(\mathbf{r}) \int \left[\iint \frac{C_{\mathbf{r}'',\mathbf{r}'''}^{b,(0)}(0)}{|\mathbf{r}-\mathbf{r}''||\mathbf{r}'-\mathbf{r}'''|}\,\text{d}\mathbf{r}''\,\text{d}\mathbf{r}'''\,\right]\notag\\
&\times 
 \biggl(\langle \hat{\rho}_{\subm{A}}(\mathbf{r}') \rangle_{0_{\subm{A}}} + \sum_{b'\in\subm{B}\backslash b}\left[\langle \hat{\rho}_{{b'}}(\mathbf{r}')\rangle_{0_{{b'}}}^{(0)} +\langle \hat{\rho}_{{b'}}(\mathbf{r}')\rangle_{0_{{b'}}}^{(1)}\right] \biggl)\text{d}\mathbf{r}\,\text{d}\mathbf{r}'\ \text{.}
 \label{eq:induction_operator2}
\end{align}
We begin by considering the expression provided by the first equality. By introducing the multipole-expanded form of the 
electrostatic potential operator of subsystem $b$ given in Eq.~\eqref{eq:pot_operator_expanded}, this becomes  
\begin{align}
\hat{\mathcal{V}}^{\text{ind}}&= \sum_{b\in\subt{B}}\int\hat{\rho}^{\text{e}}_{\subm{A}}(\mathbf{r})\langle\hat{\mathcal{V}}_{b}(\mathbf{r})\rangle_{0_{b}}^{(1)}\,\text{d}\mathbf{r}\notag\\
&=\sum_{b\in\subt{B}}\sum_{|k|=1}^{\infty}\frac{(-1)^{|k|}}{k!}\langle\hat{M}_{b}^{(k)}(\mathbf{R}_{b})\rangle_{0_{b}}^{(1)}\int\hat{\rho}^{\text{e}}_{\subm{A}}(\mathbf{r})T^{(k)}_{b\subt{r}}\,\text{d}\mathbf{r}\notag\\
&=\sum_{b\in\subt{B}}\sum_{|k|=1}^{\infty}\frac{1}{k!}\bar{{M}}_{b}^{(k)}\int\hat{\rho}^{\text{e}}_{\subm{A}}(\mathbf{r})T^{(k)}_{\subt{r}b}\,\text{d}\mathbf{r}\ \text{,}\label{eq:firstterm}
\end{align}
where we have used the definition of the induced multipole moments in Eq.~\eqref{eq:induced_multipole_moments}. Note that the multi-index summation excludes zero, since there is no induced monopole.
We identify $\hat{\mathcal{V}}_{\subm{A}}^{(k)}(\textbf{R}_{b}){=}\int \hat{\rho}_{\subm{A}}(\textbf{r})T_{\subt{r}{b}}^{(k)} \,\text{d}\textbf{r}$ as a component of the $k$'th-order derivative of the potential operator, whose expectation value gives the $k$'th-order derivative of the electrostatic potential at $\textbf{R}_{b}$, generated by the charge density of subsystem $\subt{A}$. Since it is common to work in terms of the fields,~$\hat{F}_{\subm{A}}^{(k)}(\textbf{R}_{b}){=}{-}\hat{\mathcal{V}}_{\subm{A}}^{(k)}(\textbf{R}_{b})$ for $|l|{=}1$, and its derivatives, we shall multiply Eq.~\eqref{eq:firstterm} by $1{=}({-}1)^2$, associating a minus with the interaction operator. Hereby, we arrive at the first of the two alternative multipole-expanded representations of the induction operator given by Eq.~\eqref{eq:induction_operator_expanded_2}
\begin{align}\label{firstalternative}
\hat{\mathcal{V}}^{\text{ind}}&=-\sum_{b\in\subt{B}}\sum_{|k|=1}^{\infty}\frac{1}{k!}\hat{F}_{\subm{A}}^{\text{e},(k)}(\mathbf{R}_{b})\bar{{M}}_{b}^{(k)}\ \text{,}
\end{align}
where the field and field derivative operators are defined in Eq.~\eqref{eq:field_operator}.

Proceeding to Eq.~\eqref{eq:induction_operator2} we start by rewriting 
the double integral in square brackets by introducing a Taylor expansion of the two interaction operators around a point $\mathbf{R}_b$ inside the charge distribution of subsystem $b$
\begin{align}\label{eq:C_expanded}
{\iint}\frac{C_{\mathbf{r}'',\mathbf{r}'''}^{b,(0)}(0)}{|\textbf{r}-\textbf{r}''||\textbf{r}'-\textbf{r}'''|}\, \text{d}\textbf{r}''\, \text{d}\textbf{r}''' =&\sum_{|k|=1}^{\infty}\sum_{|l|=1}^{\infty}\frac{(-1)^{|k|+|l|}}{k!\cdot l!}T_{b\subt{r}}^{(k)}P_{b}^{(k,l)}T_{b\subt{r}'}^{(l)}\notag\\[0.1in]
=& \sum_{|k|=1}^{\infty}\sum_{|l|=1}^{\infty}\frac{1}{k!\cdot l!}T_{\subt{r}b}^{(k)}P_{b}^{(k,l)}T_{\subt{r}'b}^{(l)}\ ,
\end{align}
where the last equality follows from the symmetry of the interaction tensors. $P_{b}^{(k,l)}$ is the generalized electronic polarizability defined in Eq.~\eqref{eq:responsefunction_expanded}. The summations exclude
zero, since $P_{b}^{(k,l)}$ involves transition matrix elements, which vanish for the constant monopole operator as a consequence of the orthogonality of the unperturbed wave functions of subsystem $b$. For the same reason,  $P_{b}^{(k,l)}$ contains no nuclear contribution. 
Substituting Eq.~\eqref{eq:C_expanded} into \eqref{eq:induction_operator2} yields
\begin{align}
\hat{\mathcal{V}}^{\text{ind}} =&-\sum_{b\in\subm{B}} \sum_{|k|=1}^{\infty}\sum_{|l|=1}^{\infty}\frac{1}{k!\cdot l!}\int \hat{\rho}^{\text{e}}_{\subm{A}}(\textbf{r}) T_{\subt{r}b}^{(k)}\, \text{d} \textbf{r}\ P_{b}^{(k,l)}\notag\\
& \times\int T_{\subt{r}'b}^{(l)}\biggl(\langle \hat{\rho}_{\subm{A}}(\textbf{r}')\rangle_{0_{\subm{A}}} + \sum_{b'\in\subt{B}\backslash b}\left[\langle \hat{\rho}_{b'}(\textbf{r}')\rangle_{0_{b'}}^{(0)}+\langle \hat{\rho}_{b'}(\textbf{r}')\rangle_{0_{b'}}^{(1)} \right]\biggl)\,\text{d}\textbf{r}'\ \text{.}\label{app_Vind2}
\end{align}
As before, we recognize $\hat{\mathcal{V}}_{b'}^{(l)}(\textbf{R}_{b}){=}\int T_{\subt{r}{b}}^{(l)} \hat{\rho}_{b'}(\textbf{r})\,\text{d}\textbf{r}$ as a component of the $l$'th-order derivative of the potential operator, 
which can be translated to the corresponding field and field derivative operators upon multiplication of Eq.~\eqref{app_Vind2} by $1{=}({-}1)^2$. In this case, we associate a minus with each of the interaction tensors.
Equation~\eqref{app_Vind2} can then be rewritten as
\begin{align}
\hat{\mathcal{V}}^{\text{ind}} =&-\sum_{b\in\subm{B}} \sum_{|k|=1}^{\infty}\sum_{|l|=1}^{\infty}\frac{1}{k!l!} \hat{F}_{\subm{A}}^{\text{e},(k)}(\textbf{R}_{b})  P_{b}^{(k,l)}\notag\\
&\times\biggl(\langle \hat{F}^{(l)}_{\subm{A}}(\textbf{R}_{b})\rangle_{0_{\subm{A}}} + \sum_{{b}'\in\subt{B}\backslash b}'
\left[\langle \hat{F}_{b'}^{(l)}(\textbf{R}_{b})\rangle_{0_{b'}}^{(0)}+\langle \hat{F}^{(l)}_{{b'}}(\textbf{R}_{b})\rangle_{0_{b'}}^{(1)} \right]\biggl)\ \text{.}\label{app_Vind3}
\end{align}
We shall represent $\langle \hat{F}_{{b'}}(\textbf{R}_{b})\rangle_{0_{b'}}^{(0)}$ and $\langle \hat{F}_{{b'}}(\textbf{R}_{b})\rangle_{0_{b'}}^{(1)}$ in terms of the permanent and induced multipole moments. By taking minus the $l$'th-order derivative of the expectation value of Eq.~\eqref{eq:pot_operator_expanded}, we obtain 
\begin{align}\label{app_field}
\langle \hat{F}_{b'}^{(l)}(\textbf{R}_{b})\rangle_{0_{b'}}^{(0)}=\sum_{|k|=0}^{\infty}\frac{(-1)^{|k|+1}}{k!}T_{{b'b}}^{(k+l)}M_{b'}^{(k)}\ \text{.}
\end{align}  
Analogously, the first-order correction can be recast in a multipole-expanded form as
\begin{align}
\langle \hat{F}^{(l)}_{b'}(\textbf{R}_{b})\rangle_{0_{b'}}^{(1)}=& -\int T_{\subt{r}{b}}^{(l)}\langle \hat{\rho}_{{b'}}(\textbf{r})\rangle_{0_{b'}}^{(1)}\, \text{d}\textbf{r}\notag\\
=&-\sum_{|k|=1}^{\infty}\frac{(-1)^{|k|}}{k!}T_{{b'}{b}}^{(k+l)}\int(\textbf{r}-\textbf{R}_{b})^k\langle \hat{\rho}_{{b'}}(\textbf{r})\rangle_{0_{b'}}^{(1)}\, \text{d}\textbf{r}\notag\\
=&\sum_{|k|=1}^{\infty}\frac{(-1)^{|k|+1}}{k!}T_{{b'}{b}}^{(k+l)}\bar{M}_{{b'}}^{(k)}\ \text{.}\label{app_field_inducedmoments2}
\end{align}
Finally, by substituting Eqs.~\eqref{app_field} and \eqref{app_field_inducedmoments2} into Eq.~\eqref{app_Vind3}, we arrive at the multipole-expanded representation of the induction operator given by Eq.~\eqref{eq:induction_operator_expanded}
\begin{align}
\hat{\mathcal{V}}^{\text{ind}} =&-\sum_{b\in\subm{B}} \sum_{|k|=1}^{\infty}\sum_{|l|=1}^{\infty}\frac{1}{k!\cdot l!} \hat{F}_{\subm{A}}^{\text{e},(k)}(\textbf{R}_{b})  P_{b}^{(k,l)}\notag\\
&\times\biggl(\langle \hat{F}^{(l)}_{\subm{A}}(\textbf{R}_{b})\rangle_{0_{\subm{A}}} + \sum_{{b'}\in\subt{B}\backslash{b}}\biggl[\langle \hat{F}_{{b'}}^{(l)}(\textbf{R}_{b})\rangle_{0_{b'}}^{(0)}+\sum_{|m|=1}^{\infty}\frac{(-1)^{|m|+1}}{m!}T_{{b'}{b}}^{(m+l)}\bar{M}_{{b'}}^{(m)} \biggl]\biggl)\ \text{.}\label{app_Vind4}
\end{align}

%

\end{document}